\documentclass[conference]{IEEEtran}
\usepackage{cite} 
\usepackage{tikz} 
\usepackage{xcolor}
\usepackage{url} 
\usepackage[font=small]{caption}
\usepackage{subcaption} 
\usepackage{comment} 
\usepackage{paralist}
\usepackage[frozencache,cachedir=.]{minted}
\usepackage{listings}
\usepackage{amsthm}
\usepackage{amsmath,amssymb,amsfonts}
\usepackage{algorithmic}
\usepackage{graphicx}
\usepackage{textcomp}
\usepackage{xcolor}
\usepackage{xspace}
\usepackage{pifont}
\usepackage{hyperref}
\usepackage[most]{tcolorbox}
\newtcolorbox{highlighted}{colback=yellow,coltext=black,breakable}
\usepackage{fancyhdr}

\theoremstyle{definition}
\newtheorem*{defn}{Security Definition}
\AtBeginEnvironment{minted}{%
  }

\newcommand{\para}[1]{\vspace{.03in} \noindent \textbf{#1}}
\newcommand{\subpara}[1]{\vspace{.03in} \textit{#1}}

\newcommand{\cmark}{\ding{51}}
\newcommand{\xmark}{\ding{55}}

\fancypagestyle{arxiv}
{
    \fancyhead[C]{Appears in the proceedings of the 28th Network and Distributed System Security Symposium (NDSS), 2021}
}

\ifCLASSINFOpdf
\else
\fi
\begin{document}
%
\title{DOVE: A Data-Oblivious Virtual Environment}

\author{\IEEEauthorblockN{Hyun Bin Lee\IEEEauthorrefmark{1},
Tushar M. Jois\IEEEauthorrefmark{2},
Christopher W. Fletcher\IEEEauthorrefmark{1}, and  
Carl A. Gunter\IEEEauthorrefmark{1}}
\IEEEauthorblockA{\IEEEauthorrefmark{1}University of Illinois at Urbana-Champaign:\\
\{lee559, cwfletch, cgunter\}@illinois.edu}
\IEEEauthorblockA{\IEEEauthorrefmark{2}Johns Hopkins University: jois@cs.jhu.edu\\}}


%

\ifdefined\isconf 
    \IEEEoverridecommandlockouts
    \makeatletter\def\@IEEEpubidpullup{6.5\baselineskip}\makeatother
    \IEEEpubid{\parbox{\columnwidth}{
        Network and Distributed Systems Security (NDSS) Symposium 2021\\
        21-24 February 2021, San Diego, CA, USA\\
        ISBN 1-891562-66-5\\
        https://dx.doi.org/10.14722/ndss.2021.23056\\
        www.ndss-symposium.org
    }
    \hspace{\columnsep}\makebox[\columnwidth]{}}
\else 
    \IEEEspecialpapernotice{Extended Version}
\fi 

\maketitle

\ifdefined\isconf 
\else 
\thispagestyle{arxiv}
\fi 

\begin{abstract}
Users can improve the security of remote communications by using Trusted Execution Environments (TEEs) to protect against direct introspection and tampering of sensitive data. 
This can even be done with applications coded in high-level languages with complex programming stacks such as R, Python, and Ruby. However, this creates a trade-off between programming convenience versus the risk of attacks using microarchitectural side channels. 

In this paper, we argue that it is possible to address this problem for important applications by instrumenting a complex programming environment (like R) to produce a \emph{Data-Oblivious Transcript (DOT)} that is explicitly designed to support computation that excludes side channels. Such a transcript is then evaluated on a Trusted Execution Environment (TEE) containing the sensitive data using a small trusted computing base called the \emph{Data-Oblivious Virtual Environment (DOVE)}.

To motivate the problem, we demonstrate a number of subtle side-channel vulnerabilities in the R language. We then provide an illustrative design and implementation of DOVE for R, creating the first side-channel resistant R programming stack. We demonstrate that the two-phase architecture provided by DOT generation and DOVE evaluation can provide practical support for complex programming languages with usable performance and high security assurances against side channels.
\end{abstract}
\section{Introduction}
\label{sec:intro}

Recent commercially-available Trusted Execution Environments (TEEs) such as Intel SGX~\cite{intel2014sgx,sgx-explained} and ARM TrustZone~\cite{trustzone} 
have enabled significant progress towards the outsourcing of secure computation. 
Consider for example three competing drug companies investigating genomic factors for bipolar disorder.
These companies would like to share  
their proprietary genome data and run a controlled study that releases only agreed-upon information to the three participants.
TEEs enable such use cases, without requiring trust in remote administrator software stacks such as operating systems, using a combination of hardware-level isolation and cryptographic mechanisms.

The long-term vision pursued by TEE-based software systems (e.g., \cite{haven,graphene_sgx}) is to bring TEE-level security to the masses where it can be used by data scientists familiar with existing high-level languages such as R, Ruby, and Python, but who may not have much background in security~\cite{CHAKRABARTI2020161}.

Here, we face a challenging problem.
To achieve complete security from untrusted software, it is well known that TEE software must be hardened to block a plethora of microarchitectural side channels~(e.g., \cite{sgx_grand_exposure,microscope,controlled_sc,leaky_cauldron}).
Yet, existing software-based techniques to block these channels---coming from a rich line of research in data-oblivious/constant-time programming~\cite{pc_model,practical_doprogramming,Raccoon,curve25519}---fall short of protecting existing high-level language stacks such as R, Ruby and Python.
Specifically, these techniques typically require experts to manually code core routines~\cite{curve25519,poly1305}, require the use of custom domain-specific languages~\cite{FactLanguage, SGXBigMatrixDOVE}, or only apply to close-to-metal compiled languages~\cite{pc_model,Raccoon}. 
Modern high-level languages, however, require complex stacks to support interpreted execution, just-in-time compilation, etc.
As a case-in-point, the popular R stack features almost a million lines of code written in a combination of C, Fortran, and R itself~\cite{r_core}.
Subtle issues in any of this code create security holes. 

The goal of this paper is to extend 
data-oblivious/constant-time techniques 
to apply to existing high-level, interpreted languages, thus enabling TEE-level security for non-experts.
The key strategy and insight is this: 
\emph{if key observable features of a computation are truly independent of sensitive data, then that computation can be carried out with a collection of stand-ins for the data}. 
We call these stand-ins ``pseudonyms''.

To exploit this idea we perform computation in two phases. 
In the first phase, we run the target computation on pseudonyms in the chosen high-level language, like R or Python.
Since there is no sensitive data present, this stage cannot leak sensitive information. 
We instrument the programming stack so that this evaluation on pseudonyms outputs what we call a ``Data-Oblivious Transcript (DOT)''.
The DOT is akin to a straight-line code representation of the original program, i.e., the transcript of operations performed when the program is evaluated on the pseudonyms.   
In the second phase of our computation, we evaluate the DOT on a small Trusted Computing Base (TCB) that runs within a TEE.
This TEE contains the sensitive data, which is used in place of the pseudonyms.
Protecting sensitive data \emph{after} the DOT is constructed is relatively straightforward.
Since the DOT is similar to straight-line code, the TEE need only apply simple transformations to evaluate it in a data-oblivious fashion on real hardware.
In the worst case, where the original computation was actually data dependent on the pseudonyms, the resulting computation in the TEE may be functionally incorrect but leaks no sensitive information.

Conceptually, the DOT plays a role similar to a compiler intermediate representation.
Our approach can be characterized as a \emph{frontend} translating high-level evaluation to the DOT and a \emph{backend} evaluating the DOT data-obliviously, on sensitive data.
A key benefit of this decoupled approach is that only the backend (importantly, \emph{not} the frontend) is part of the TCB.
This provides a powerful strategy for protecting complex, high-level programming stacks against side channel attacks.
In addition to a reduced TCB, the decoupling provides modularity and extensibility benefits similar to those found in modern compilers.
For example, to add support for a new high-level language, we need only change frontend code.
Likewise if a security vulnerability is found in the TEE, or we wish to deploy different TEEs to protect execution for different processor microarchitectures, we need only change backend code.

Putting it all together, we design and implement an instance of the above architecture, called the ``Data-Oblivious Virtual Environment (DOVE)''. 
As a proof-of-concept, we develop a DOVE frontend that translates programs written in the R language to a DOT representation 
and design a backend that evaluates the DOT on sensitive data inside of an Intel SGX enclave.  

To validate DOVE, we show how to support a third-party library of genomics analysis algorithms written in R~\cite{eva_chan}, which we call the \textit{evaluation programs}. 
Out of 13 evaluation programs, DOVE can run 11 of them, with these 11 totaling 326 lines of R code. 
For 10 of the 11 above programs, our frontend can automatically convert the unmodified R program into the DOT language; converting the remaining case required manual user intervention because of a programming construct not yet supported by our frontend.
We collect performance benchmarks on these programs with a real-world genomic dataset consisting of three populations of honeybees~\cite{bee_study}.

\para{Summary of contributions.}
\begin{compactenum}
    \item We identify a number of subtle side-channel vulnerabilities in the R language. 
    \item We design DOVE, the first architecture that runs existing high-level interpreted languages and is demonstrably resistant to side channels. 
    \item We provide an implementation of DOVE for R, creating the first side-channel resistant R programming stack.
    \item We evaluate the security and performance of DOVE against evaluation programs drawn from the genomics literature. Relative runtime overheads of DOVE against vanilla R on these programs range from \(12.74\times\) to \(341.62\times\).
\end{compactenum}

Source code for the DOVE frontend and backend prototype is available at \url{https://github.com/dove-project}.

Finally, 
the extended version of this paper~\cite{dove_full} additionally includes
1) a grammar for the DOT language, 
2) more details on the evaluation programs and 
3) more details regarding the use of the Intel Performance Counter Monitor (PCM) APIs for our security evaluation.
\section{Background}
\label{sec:background}
\subsection{Programming in R}
\label{sec:background:R}
R is a statistical language that provides convenient interfaces for computations on arrays and matrices. Most function calls including primitive operators like addition and subtraction perform element-wise operations on array-like values. 
Figure~\ref{fig:r_snippet} is an R code example from our evaluation programs that includes such operations.

\para{Computation in R.} 
R is an interpreted language~\cite{r_core}, and its interpreter is written mostly in C and to a lesser extent Fortran and R itself. 
Every object is represented with a symbolic expression (S-expression)~\cite{sexp} such that interpreter parses R statements into S-expressions. The S-expressions are then evaluated and dispatched to the corresponding library functions written in C. Each C function runs on hardware as a compiled binary object. Thus, analyzing code written in R is more complex than analyzing code that is directly compiled and run on hardware (e.g. C, C++).

\para{Not Applicable (NA).}  
R represents \mintinline{c}{null}-like, empty values with \mintinline{s}{NA}, the representation of which depends on the datatype. A real-valued S-expression in R is represented with a IEEE 754 \mintinline{c}{double}; \mintinline{c}{NA_REAL} is defined with the special double value \mintinline{s}{NaN} with a specific lower word (\mintinline{c}{1954}). 
The interpreter treats \mintinline{s}{NA} differently from other values, even from \mintinline{s}{NaN}.
Integer and logical (i.e., boolean) S-expressions are implemented with an \mintinline{c}{int} type, so 
R reserves the lowest integer value \mintinline{c}{INT_MIN} for the representation of \mintinline{c}{NA_INTEGER} and \mintinline{c}{NA_LOGICAL}. 

\para{S3 method dispatch.}
The most common object-oriented programming system in R is S3 method dispatch.
For each function call on an object, the S3 object system calls the correct method associated with that object. For example, for \mintinline{s}{print(x)}, when \mintinline{s}{x} is a scalar, S3 calls \mintinline{s}{print.numeric(x)}; when \mintinline{s}{x} is a matrix, S3 calls \mintinline{s}{print.matrix(x)} instead. 
A programmer who wishes to add their custom type \mintinline{s}{myObject} to \mintinline{s}{print} would define a function \mintinline{s}{print.myObject(x)}. This paradigm makes it easy to supply new types of objects to existing functions, making the differences in implementation transparent to the end user. S3 is the OOP system used in the base R, making it especially useful to override commonly used functions.

\subsection{Microarchitectural Side-Channel Attacks}
\label{sec:background:side_channels}

Microarchitectural (shortened as ``\(\mu Arch\)'') side-channel attacks are a class of privacy-related vulnerabilities where a sensitive program's hardware resource usage leaks sensitive information to an adversary co-located to the same (or a nearby) physical machine~\cite{sok_side_channels}.
Over the years, numerous hardware structures---a variety of cache architectures~\cite{cache_sc,flush+,cache_bleed,noninclusive}, branch predictors~\cite{branchpred_sc,BranchScope}, pipeline components~\cite{FPU_leaky,Mult_leaky,portsmash} and other structures~\cite{4kalias,controlled_sc,leaky_cauldron,tlb_bleed,drama,rand_ccs16}---have been found to leak information in this way.
Many of these attacks require that the attacker only share physical resources with the victim (e.g., Prime+Probe and the cache~\cite{last_level_cache_practical,cache_sc} or Drama and the DRAM row buffer~\cite{drama}), as opposed to sharing virtual memory with the victim (e.g. \cite{flush+}).


\subsection{Enclave Execution and Intel SGX}
\label{sec:background:sgx}
Enclave execution~\cite{EnclaveCCS17}, such as with Intel SGX~\cite{intel2014sgx}, protects sensitive applications from direct inspection or tampering from supervisor software.
That is, the OS, hypervisor and other software are considered to be the attacker~\cite{ZeroTrace,oisa,OblivML,Raccoon,sgx_cache_attacks,sgx_single_step_cache_attack,sgx_grand_exposure,leaky_cauldron,cache_zoom,microscope}, who will be referred to as the \emph{SGX adversary} for the rest of the paper.
To use SGX, users partition their applications into enclaves at some interface boundary.
For example, prior work has shown how to run whole applications with a LibOS~\cite{haven,graphene_sgx}, containers~\cite{panoply}, and data structure abstractions~\cite{ZeroTrace} within enclaves.
At boot, hardware uses attestation via digital signatures to verify the user's expected program and input data are loaded correctly into each enclave.
Isolation mechanisms implemented in virtual memory protect enclave integrity and confidentiality during execution.

SGX uses the Enclave Page Cache (EPC) to store enclave application code and data. The EPC is stored in a protected region of memory known as Processor-Reserved Memory (PRM). The processor prevents other system components from reading the PRM with the help of another component, the Memory Encryption Engine (MEE), that provides encryption and integrity protection for the PRM~\cite{mckeen2013innovative}. The EPC has a fixed size of 64 or 128 MB, shared among all enclaves~\cite{intel2020sgxdev}.
For applications requiring more memory,
SGX uses an EPC paging mechanism supported by the SGX OS driver. 
Specifically, the OS can move pages out of/into the EPC and manipulate them as if they were regular pages from a demand-paging perspective.
For security, pages moved out of/into the EPC are transparently encrypted/decrypted and integrity checked by the SGX hardware~\cite{intel2014sgx, mckeen2013innovative}.

\para{Side-channel amplification.}
Despite providing strong virtual isolation, SGX enclave code is still managed by untrusted software.
Prior work has shown how this exacerbates the side-channel problem described in Section~\ref{sec:background:side_channels}.

First, SGX does not provide any physical isolation.
Thus, nearly all of the \(\mu Arch\) side-channel attacks discussed in Section~\ref{sec:background:side_channels} immediately apply in the SGX setting.

Second, importantly, the OS-level attacker has significant control over the enclave's execution and the processor hardware and thus can orchestrate finer-grain, lower-noise attacks than would otherwise be possible.
For example, controlled side-channel attacks~\cite{controlled_sc} and follow-on work~\cite{leaky_cauldron} provide a zero-noise mechanism for an attacker to learn a victim's memory access pattern at page (or sometimes finer) granularity.
A line of work has further shown how the attacker can effectively single-step, and even replay, the victim to measure fine-grain information such as cache access pattern and arithmetic unit port contention~\cite{sgx_cache_attacks,sgx_grand_exposure,cache_games,cache_zoom,sgx_single_step_cache_attack,sgxstep,microscope}.

\subsection{Data-Oblivious Programming}
\label{sec:background:data_oblivious_programming}

Data-oblivious (sometimes called ``constant-time'' in the hardware setting) programming is a way to write programs that makes program behavior independent of sensitive data, with respect to the side channels discussed in Section~\ref{sec:background:side_channels}~\cite{curve25519, poly1305, pc_model,Raccoon,IronDOVE,OblivML,SGXBigMatrixDOVE,Opaque,OblivSQL,Oblix,ObliviousUtils,ZeroTrace,obliviate,FPU_leaky,practical_doprogramming,Lobliv,FactLanguage,Oblivm,ZahurCompiler,path_oram, OBFS, GraphSC, OStableMatching, CacheObliviousSort, Oblivm,tiny_garble,ZahurOakland,ODS}.
In the hardware setting, what constitutes data-oblivious execution depends on the intended adversary.
In the SGX setting, we must assume a powerful adversary that can monitor potentially any \(\mu Arch\) side channel as described in Section~\ref{sec:background:sgx}.

Thus, prior works that try to achieve data obliviousness in an SGX context~\cite{Raccoon,IronDOVE,OblivML,SGXBigMatrixDOVE,Opaque,OblivSQL,Oblix,ZeroTrace,obliviate} implement computation using only a carefully chosen subset of arithmetic operations (e.g., bitwise operations), conditional moves, branches with data-independent outcomes, jumps with non-sensitive destinations, and memory instructions with data-independent addresses.
For example, an \mintinline{s}{if} statement with a sensitive predicate is implemented as straight line code that executes both sides of the \mintinline{s}{if} and uses a data-oblivious ternary operator (such as the x86 \mintinline{nasm}{cmov} instruction or the CSWAP operation) to choose which result to keep.

\section{Threat Model}
\label{sec:threat_model}

In this paper we consider a setting where one or more users submit data to an untrusted server that computes on said data in a high-level language such as R.
The server hosts SGX as well as a regular software stack outside of SGX.
The user(s) and SGX hardware mechanism are trusted.
The program computing on user data, like the R interpreter and evaluation programs, is assumed to be non-sensitive.
No software running on the remote host outside of an SGX enclave is trusted---this includes the supervisor software stack, disks, the connection between client and server, and the other hardware components besides the processor hosting SGX.
Per the usual SGX threat model (Section~\ref{sec:background:sgx}), we assume the OS is compromised and may run concurrently on the same hardware, such as in adjacent hyperthread/SMT contexts, on neighboring physical cores, etc.

\para{Security goal.}
Our goal is to prevent arbitrary non-SGX enclave software from learning anything about the users' data, other than non-sensitive information about the data such as its bit length.
Given SGX's architecture, this implies protecting user data from leaking over arbitrary non-speculative \(\mu Arch\) side channels (Section~\ref{sec:background:side_channels}), given the powerful SGX adversary described in Section~\ref{sec:background:sgx}. This is formalized in our security analysis (Section~\ref{sec:sec_analysis}).

\para{Security non-goals.}
We do not defend against hardware attacks such as power analysis~\cite{KocherDPA}, EM emissions~\cite{eddie}, compromised manufacturing (\textit{e.g.}, hardware trojans~\cite{a2}), or denial of service attacks.
Also, our current implementation does not have mechanisms to mitigate speculative execution attacks~\cite{spectre} beyond default SGX protections (e.g., flushing branch predictor state on context switches~\cite{SgxPectre}).
If additional protection is needed, our backend (an SGX enclave) can be re-compiled with a software-level protection such as speculative load hardening~\cite{slh}.

\para{Functional correctness.}
While we designed DOVE to preserve semantic equivalence (functional correctness) between the input high-level program and output DOT plus its subsequent execution, we do not have a formal proof that our implementation does indeed preserve semantic equivalence.
This is in line with other data-oblivious compilation frameworks (e.g., \cite{Oblivm}) and we consider such end-to-end verified compilation to be important future work.
Note that even if the DOVE frontend has bugs, leading to functionality- or security-related issues, our security guarantee in Section~\ref{sec:sec_analysis} still holds. 

Note, when we refer to \emph{trusted computing base} (TCB) we mean the DOVE software that must function as intended---i.e., be free of logic bugs and control-flow hijacking vulnerabilities---for security to hold.
\section{Attack Examples}
\label{sec: attack_examples}

A major problem this paper addresses is how to protect R programs from the SGX adversary.
As a starting point, imagine we try to run secure R code by moving the whole R stack into the SGX enclave (which is the approach taken by prior work~\cite{haven,graphene_sgx}).
We demonstrate subtle \(\mu Arch\) side-channel attack vectors that come up in this approach, using the code snippet in Figure~\ref{fig:r_snippet} as a guiding example. 
This code is found in 4 of 13 evaluation programs that form a public repository of code for genomics research. 
Three additional programs from these evaluation programs feature a similar snippet. We explain what these programs are in Section~\ref{sec: perf_eval: eval_scripts}.
The program takes as input a set of samples made up of diploid Single Nucleotide Polymorphisms (SNP) sequences and outputs the number of samples that express a given genotype for each SNP position. 
We use R version 3.2.3 to illustrate these attack examples.

\subsection{Example Walkthrough}
The program represents the database of samples as \mintinline{s}{geno}, an $m$ by $n$ matrix, where each column is one of $n$ samples, each of which has $m$ SNP positions.
Each position in the matrix has a genotype, denoted as an integer \mintinline{s}{0}, \mintinline{s}{1} or \mintinline{s}{2}.
The sensitive data is the contents of \mintinline{s}{geno}, namely which genotype each SNP is for each sample.
The matrix dimensions ($m$ and $n$) are non-sensitive.

Computationally, the code works as follows.
Line~\ref{line:bwand} sanitizes the input database: any entry that is not one of the three allowed genotypes is replaced with the special value \mintinline{s}{NA} (Section~\ref{sec:background:R}).
This occurs in real data due to noise in the sequencing process; in particular, 1.5\% of the SNP entries in the honeybee dataset that we use in Section~\ref{sec: perf_eval: eval_scripts} are marked as \mintinline{s}{NA}.
The code first computes element-wise filters \mintinline{s}{geno != 0}, \mintinline{s}{geno != 1}, \mintinline{s}{geno != 2}, each of which produces a matrix of booleans (a mask) indicating whether the condition is satisfied for each SNP position in each sample.
The logical AND (\mintinline{s}{&}) performs element-wise AND of these 3 masks (producing a new mask) which is used to conditionally assign elements in \mintinline{s}{geno} to \mintinline{s}{NA}. 
Then, the code in Lines~\ref{line:loop:start}--\ref{line:loop:end} produces three vectors 
\mintinline{s}{n0,n1,n2}, where R ``applies'' the \mintinline{s}{sum()} function on each row (specified by the second argument \mintinline{s}{1}) such that each vector is the count (sum on rows) of the number of samples that express each genotype.

Given the above code, the adversary's goal is to learn the genotype at each SNP position---that is, whether the value of each cell in \texttt{geno} is \mintinline{s}{0,1,2} or \mintinline{s}{NA}.
Importantly, given no additional information about R's implementation, \emph{the R-level code in Figure~\ref{fig:r_snippet} follows guidelines for achieving data-obliviousness} (Section~\ref{sec:background:data_oblivious_programming}), which would seemingly prevent leaking the above information.
For example, it applies simple arithmetic/logical operations element-wise over matrices of non-sensitive size, performs a count over a subset of samples with a non-sensitive length, etc. 
Yet, as we now show, this code nonetheless leaks privacy through \(\mu Arch\) side channels.


\begin{figure} 
\centering
\caption{R code snippet. \mintinline{s}{geno} is a sensitive diploid dataset.}
\footnotesize
\begin{minted}[xleftmargin=20pt,linenos,breaklines,breaksymbolleft={},breakindent=12pt,escapeinside=||]{s}
geno[(geno!=0) & (geno!=1) & (geno!=2)] <- NA |\label{line:bwand}|
geno <- as.matrix(geno)
n0 <- apply(geno==0,1,sum,na.rm=T)|\label{line:loop:start}|
n1 <- apply(geno==1,1,sum,na.rm=T)|\label{line:loop:mid}|
n2 <- apply(geno==2,1,sum,na.rm=T)|\label{line:loop:end}|
\end{minted}
\label{fig:r_snippet}
\end{figure}

\subsection{Logical Operators}
\label{sec: attack_examples: bitwise_and}

We start with Line~\ref{line:bwand} in Figure~\ref{fig:r_snippet}, specifically the logical \mintinline{s}{&} operations performed between the masks.
At the level of R code, these look like safe data-oblivious operations (Section~\ref{sec:background:data_oblivious_programming}).
Recall that the dimensions of \mintinline{s}{geno} are non-sensitive.
Thus, combining each mask with \mintinline{s}{&} entails performing a data-independent number of simple logical operations (\mintinline{s}{&}); this is traditionally regarded as safe. 

\textit{Yet, this code is not data-oblivious thanks to the transformations it undergoes in the R stack before reaching hardware.} 

First, the code is transformed from R into C calls by the R interpreter, shown in Figure~\ref{fig:interpret_and_compile:interpret}. 
When R interprets \mintinline{s}{&}, it invokes the C routine given in Figure~\ref{fig:interpret_and_compile:interpret}.
This snippet takes different code paths, depending on the values of \mintinline{s}{x1} and \mintinline{s}{x2}, which the SGX adversary can detect by single-stepping~\cite{sgxstep} or by replaying the victim~\cite{microscope} and measuring time, branch predictor state, etc. (see below).
In this case, the attacker learns if one of \mintinline{s}{x1} or \mintinline{s}{x2} equals 0.
Since this \mintinline{s}{&} is applied to each SNP position of each sample, this information is leaked for every SNP position.

Second, the compiler compiles the resulting C into assembly, which leaks additional information.
Consider Figure~\ref{fig:interpret_and_compile:compile}, which is the assembly for Lines~\ref{line:bonz_and} to \ref{line:assignz_and} in Figure~\ref{fig:interpret_and_compile:interpret}, Note that the C standard requires short-circuit evaluation for the logical \mintinline{c}{||} operator such that if the left operand is true, the right operand is not evaluated. 
Depending on the outcome of the left predicate \mintinline{s}{x1 == 0}, the code at the assembly level will again take different paths.
Hence, the attacker learns not only whether one of \mintinline{s}{x1} or \mintinline{s}{x2} equals 0, but also learns information about \emph{which} one of them equals 0.

Figure~\ref{fig:bitwise_and_backend} counts the number of instructions executed at the assembly level for each possible input to \mintinline{s}{&}.
Confirming the above explanation, we see that the instruction count equals 45 if and only if \mintinline{s}{x1} equals 0.
Thus, the adversary learns whether this is the case if it can monitor a function of the instruction count.
Other cases leak other pieces of information such as whether both \mintinline{s}{x1} and \mintinline{s}{x2} equal 1.

To test how small differences in instruction count translate into measurable effects, we conduct a simple experiment. 
We measure the number of cycles taken to evaluate one million iterations of expression \mintinline{s}{0 & 0} against those of \mintinline{s}{1 & 0}.
Note that the execution length of these two expressions only differ by two x86-64 instructions in  Figure~\ref{fig:bitwise_and_backend}. 
Having access to a large number of measurements may occur naturally, \textit{e.g.}, if the sensitive data is accessed in a loop, or if the attacker performs a \(\mu Arch\) replay attack~\cite{microscope}.
We make 100 trials of such measurements against R with the Intel Performance Counter Monitor (PCM)~\cite{intel_pcm}. On average, it took \(\mu_{00}=\) 73.9 million cycles (\(\sigma_{00}=\) 441K) for \mintinline{s}{(0 & 0)}, but it took \(\mu_{10}=\) 75.2 million cycles (\(\sigma_{10}=\) 416K) for \mintinline{s}{(1 & 0)} on average; the cycle count differences vary by a noticeable margin in the evaluation of these two expressions. 

\begin{figure}[!htb]
\caption{The R interpreter implementation of the \mintinline{s}{&} operator.}
\begin{subfigure}[t]{\columnwidth}
\subcaption{C source code snippet of the \mintinline{s}{&} operator implementation.}
\footnotesize
\centering
\begin{minted}[xleftmargin=20pt,linenos,breaklines,breaksymbolleft={},breakindent=12pt,escapeinside=!!]{c}
if (x1 == 0 || x2 == 0) !\label{line:bonz_and}!
    pa[i] = 0; !\label{line:assignz_and}!
else if (x1 == NA_LOGICAL || x2 == NA_LOGICAL)
    pa[i] = NA_LOGICAL;
else
    pa[i] = 1;
\end{minted}
\label{fig:interpret_and_compile:interpret}
\end{subfigure}
\begin{subfigure}[t]{\columnwidth}
\subcaption{The Intel-syntax x86-64 assembly for Lines~\ref{line:bonz_and} and \ref{line:assignz_and} of the C code in Figure~\ref{fig:interpret_and_compile:interpret}, lightly edited for clarity.}
\footnotesize
\centering
\begin{minted}{nasm}
; x1 in [rbp-0x58], x2 in [rbp-0x54]
a8: cmp   DWORD PTR [rbp-0x58],0x0 ; x1==0
ac: je    b4 ; if true, jump to pa[i]=0
ae: cmp   DWORD PTR [rbp-0x54],0x0 ; x2==0
b2: jne   cf ; if false, jump to else if
b4: mov   rax,QWORD PTR [rbp-0x50]
b8: lea   rdx,[rax*4+0x0]
c0: mov   rax,QWORD PTR [rbp-0x8]
c4: add   rax,rdx ; calc addr of pa[i]
c7: mov   DWORD PTR [rax],0x0 ; pa[i]=0
cf: ...
\end{minted}
\label{fig:interpret_and_compile:compile}
\end{subfigure}

\label{fig:interpret_and_compile}
\end{figure}

\begin{figure}[!htb]
\footnotesize
\centering
\caption{The associated x86-64 instruction counts for different permutations of \mintinline{s}{x1} and \mintinline{s}{x2} fed as input to \mintinline{s}{&} in R.}
\begin{tabular}{|l|c|c|}
\hline
Expression & Value & Instruction Count \\
\hline
\mintinline{s}{0 & 0} &  \mintinline{s}{0}  & 45 \\
\mintinline{s}{0 & 1} &  \mintinline{s}{0}  & 45 \\
\mintinline{s}{1 & 0} &  \mintinline{s}{0}  & 47 \\
\mintinline{s}{1 & 1} &  \mintinline{s}{1}  & 54 \\
\mintinline{s}{0 & NA} &  \mintinline{s}{0}  & 45 \\
\mintinline{s}{1 & NA} &  \mintinline{s}{NA}  & 57 \\
\mintinline{s}{NA & 0} &  \mintinline{s}{0}  & 47 \\
\mintinline{s}{NA & 1} &  \mintinline{s}{NA}  & 53 \\
\mintinline{s}{NA & NA} &  \mintinline{s}{NA}  & 53 \\
\hline
\end{tabular}
\label{fig:bitwise_and_backend}
\end{figure}

Similar issues exist for other logical operators \mintinline{s}{|} and \mintinline{s}{xor()}.
In fact, \mintinline{s}{xor()} is implemented using R-level \mintinline{s}{&} and \mintinline{s}{|} operators.
Even binary comparison operators such as \mintinline{s}{==} and \mintinline{s}{!=} have similar issues.
For example, R's implementation of both these operators uses branches at the R level to first check if either operand is \mintinline{s}{NA}. 

\subsection{Functions}
\label{sec:attack_examples:rowsums}
Aside from R-level primitive operators, R also has a large library of functions written in either R or C.
In Figure~\ref{fig:r_snippet}, we see base R functions \mintinline{s}{as.matrix()}, \mintinline{s}{apply()}, and \mintinline{s}{sum()}. The \mintinline{s}{as.matrix()} function simply converts \mintinline{s}{geno} to a matrix object, similarly to \mintinline{c++}{dynamic_cast} in C++. 

In Line~\ref{line:loop:start},
\mintinline{s}{geno==0} produces a matrix of booleans (a mask) similar to those created on Line~\ref{line:bwand}.
Then, \mintinline{s}{apply()} invokes the \mintinline{s}{sum()} function for each row (dimension 1) that counts the occurrence of \mintinline{s}{TRUE} in each row of the argument matrix. Calls to \mintinline{s}{apply()} in Lines~\ref{line:loop:mid}-\ref{line:loop:end} have similar issues.

\mintinline{s}{sum()} for integers and booleans is implemented in C as \mintinline{c}{isum()} in the R source~\cite{r-source}. 
Interestingly, this code does perform accumulations data-obliviously; that is, each boolean is treated as 0 or 1 and accumulated without a branch that checks \mintinline{s}{if (TRUE)} or \mintinline{s}{if (FALSE)}.
Yet, the code \emph{does still branch} based on whether the current value is \mintinline{s}{NA} before accumulation, once again leaking which entries are \mintinline{s}{NA}. 

\subsection{Data-Dependent Constructs}
\label{sec:attack_examples:if}

Finally, any code construct that is not data-oblivious in C is more-than-likely not data-oblivious in R.
For example, an \mintinline{c}{if} statement in C with a sensitive predicate can reveal that predicate to the SGX adversary~\cite{branchpred_sc,BranchScope}.
Likewise, an \mintinline{s}{if} statement in R with a sensitive predicate causes an even larger (easier to measure) perturbation in program execution, due to the additional steps taken to execute that branch on hardware.

\subsection{Discussion} 
\label{sec: attack_examples: discussion}

These examples are only a small subset of the parts of R that leak sensitive information. 
R is a large code base comprising 992,564 lines of code with sophisticated runtime mechanisms such as just-in-time compilation~\cite{jitleaks},
and is composed of hundreds of API functions and other features, implemented in a combination of R, C and Fortran \cite{r_core}.\footnote{Specifically, there are 388,141 lines of C, 345,547 lines of R and 258,876 lines of Fortran in the version of the R source we used for this paper.}
Not to mention, even when an R script makes it to assembly code, we must still worry about microarchitecture-specific unsafe instructions that modulate hardware resources as a function of their input operands (e.g., \cite{FPU_leaky,Mult_leaky,practical_doprogramming,oisa}).

This presents a serious security problem.
Many data scientists and statisticians use R to compute on sensitive data every day.
Clearly, it is not tractable for these users to understand the security implications of the code they write.
At the same time, R's large code base makes manually patching data leaks inherently haphazard and error prone, even for security experts.
As a result, experts have hitherto focused on replicating R's functionality in a new language/stack \cite{SGXBigMatrixDOVE}.

In the next section, we address this challenge by designing the first secure R stack, where data scientists can program in (nearly) unchanged R, interact with the same R functionality with which they are familiar, and have strong confidence there are no latent side channels. 
\section{Design}
\label{sec: arch_lang}

We now describe the Data-Oblivious Virtual Environment (DOVE).
This begins with a design overview and summary of design benefits (Sections~\ref{sec:design:overview}, \ref{sec:design:benefits}).
Section~\ref{sec:design:dot} discusses the Data-Oblivious Transcript (DOT), which serves as the link between high-level programming and data-oblivious execution. 
Section~\ref{sec:design:frontend} discusses the DOVE frontend, which is a set of classes that convert R code into the DOT, using pseudonyms instead of sensitive data. 
Finally, Section~\ref{sec:design:backend} describes the DOVE backend, an SGX enclave that converts the DOT operations on pseudonyms to data-oblivious computation on the actual sensitive data.

\subsection{Design Overview}
\label{sec:design:overview}

As stated in the threat model (Section~\ref{sec:threat_model}), DOVE's security objective is to evaluate programs written in high-level (e.g., interpreted) languages in a data-oblivious manner.
The key insight is that an operation that is truly data oblivious does not require the actual data to be present. 
Instead, the operation can take place on a \emph{pseudonym} of the data. 
These pseudonyms have the same interface as normal data of the same type and support the same operations. 
For example, matrices are replaced with matrix pseudonyms, and matrix pseudonyms can be computed upon using the same operations as normal matrices (e.g., element-wise addition, matrix multiplication). 
However, the pseudonym contains no sensitive data, i.e., all of its data entries are replaced with \(\bot\). 
This pseudonym
is constructed solely through non-sensitive information specified for each pseudonym, such as, for matrices, the number of rows and columns. 
However, since the pseudonym does not actually have the data, any operation on the pseudonym is functionally equivalent to a NOP, i.e.,  $\ast \oplus \bot \rightarrow \bot$ where $\ast$ is a wildcard for any data value and $\oplus$ is an operation on the data.
Instead, the operation performed is appended
to a log. This log, which we call a \emph{Data-Oblivious Transcript (DOT)}, is thus akin to
a straight-line representation of the execution of the input program. 
The DOT can then be replayed on the \emph{actual} data, executing the same operations as the input program.
 
\begin{figure}[t]
    \centering
    \includegraphics[width=1\linewidth]{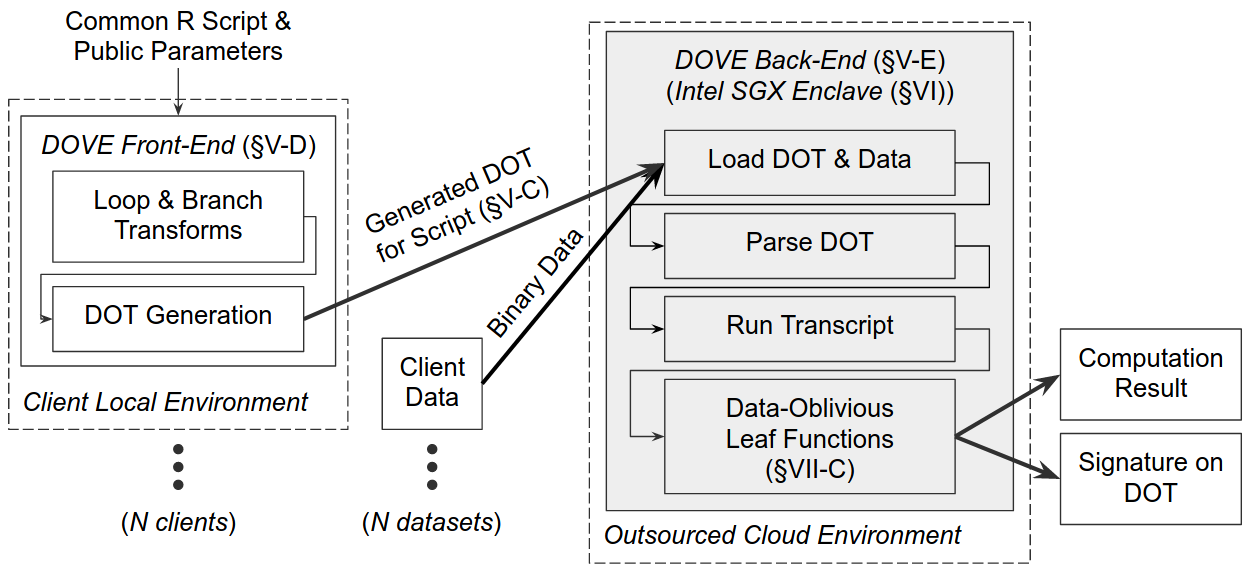}
    \caption{High-level overview of DOVE. 
    Bold-face arrows between nodes represent communication over (mutually-authenticated) TLS, while thinner ones are intra-process communication within a component.
    Shading indicates the location of our trusted computing base (TCB).}
    \label{fig:diag_arch:overview}
\end{figure}

With this in mind we propose the following architecture, shown in Figure~\ref{fig:diag_arch:overview}.
Our architecture is broken into two components, making up a \emph{frontend} and \emph{backend}.
Each of $N$ clients runs the same input --- a common (non-sensitive) high-level program --- in their local environment (``frontend''). The frontend replaces any references 
to sensitive data with pseudonyms and generates a DOT of the input program.
Although only a single DOT needs to be generated for evaluation later on, each client can optionally compute its own DOT for program integrity-checking purposes (see Section~\ref{sec:design:frontend} for more information).
This TEE (``backend'') hosts the DOVE virtual machine, which is built with data-oblivious primitives.
The virtual machine checks that all DOTs are equivalent (optional, for integrity) 
and runs the operations listed on the actual data.

Intuition for security comprises two parts.
First, because the DOT is conceptually an execution trace, the backend TEE evaluates the same operations in the same order as the R program input to the frontend, regardless of the sensitive data provided to the backend.
Importantly, the DOT was not created using any sensitive data, so the functions listed in the DOT are inherently independent/oblivious of that data. 
Second, we will architect the backend to ensure each operation is data oblivious, using well-established techniques for constant-time/data-oblivious execution.

The above architecture is general. The frontend can be adapted for different high-level languages (e.g., R, Python, Ruby), and the backend can be implemented for a variety of TEEs (e.g., SGX, TrustZone).
For the rest of the paper, we explain, design, and evaluate ideas assuming the frontend input language is R and the TEE is SGX.

\para{Sensitive \& non-sensitive Data.} 
Our goal is to make execution independent of sensitive data from a \(\mu Arch\) side channel perspective. 
However, like other systems enforcing a similar information flow policy, DOVE allows for \emph{non-sensitive} (i.e., public) data to influence attacker-visible execution. This is an important performance optimization.
For example, matrix operations on rows and columns are common in our target domain; such operations' performance is largely a function of the matrix dimensions, which are usually non-sensitive. 
DOVE performs optimizations based on non-sensitive data during DOT construction (in the frontend) by generating the DOT with concrete non-sensitive inputs.
This includes concretizing dimension arithmetic, loop bounds and control flow---when they are not a function of sensitive data.
As we discuss in Section~\ref{sec:design:frontend:disallowed}, certain constructs such as loops with sensitive loop bounds are disallowed.

\subsection{Benefits of Proposed Architecture}
\label{sec:design:benefits}

The above two-phase architecture has the following security, performance, and extensibility benefits. 

\begin{compactitem}
\item \textbf{Small trusted computing base.} 
The only part of the DOVE architecture that actually handles sensitive data is the backend, which is made up of a relatively small C/C++ codebase featuring 7,001 lines of code, 4,295 of which consist of a previously-vetted, external data-oblivious fixed point library~\cite{FPU_leaky}.
Importantly, the R stack (with its almost 1 million lines of code~\cite{r_core}) is not in the trusted computing base.

\item \textbf{No use of cryptographic encrypted computation.}
Our design performs data-oblivious computation without resorting to encrypted computation techniques (such as homomorphic encryption and garbled circuits).

\item \textbf{Minimal changes to programmer-facing interface.} 
The DOVE frontend performs a set of automated transformations (e.g., if-conversion, converting loops with early exits to guarded loops~\cite{practical_doprogramming,Oblivm}) to make input R code compatible with DOT semantics.
As we show in our evaluation, the client can typically submit their unmodified R programs to the frontend and therefore is not required to learn a new language. 

\item \textbf{Extensibility to other languages.} 
Because the DOT decouples \emph{R semantics} from \emph{backend semantics}, DOVE can in principle support additional languages by implementing a new frontend without rewriting the backend. 

\item \textbf{Extensibility to other threat models.} 
Because the DOT decouples \emph{backend semantics} from \emph{R semantics},
if a new \(\mu Arch\) side channel is discovered that undermines backend security, the backend can be patched locally without necessarily making a change to the frontend. 
\end{compactitem}

\subsection{Data-Oblivious Transcript (DOT)}
\label{sec:design:dot}

\begin{figure}[t]
    \centering
    \includegraphics[scale=0.35]{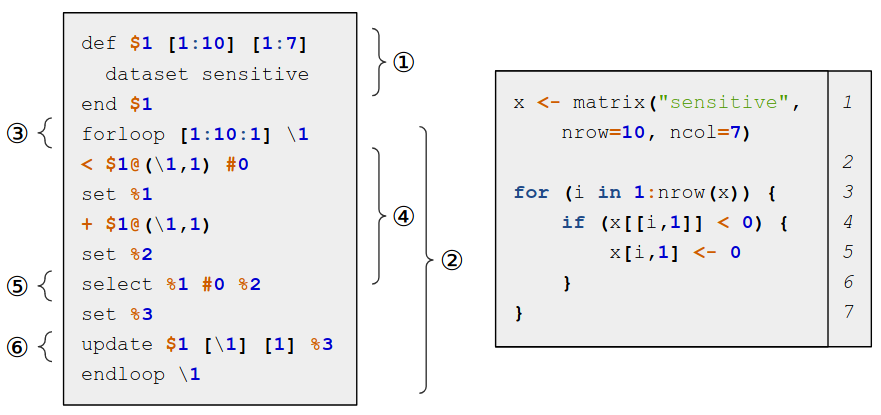}
    \caption{A DOT (left) and its associated R program (right). The matrix \mintinline{s}{x} corresponds to the pseudonym \mintinline{haskell}{$1} in the DOT, and the loop index \mintinline{s}{i} with \texttt{\textbackslash1}. \textcircled{\raisebox{-0.4pt}{1}} corresponds to line 1 of the program, \textcircled{\raisebox{-0.4pt}{2}} the \mintinline{s}{for} loop on line 3, \textcircled{\raisebox{-0.4pt}{3}} the \mintinline{s}{if} statement on line 4, and \textcircled{\raisebox{-0.4pt}{6}} the assignment in line 5. Intermediate values are stored in variables marked with \texttt{\%}, and constants are declared using \texttt{\#}. 
    }
    \label{fig:diag_arch:fend}
\end{figure}

The Data-Oblivious Transcript, or DOT, forms the core of the DOVE architecture, bridging an input program written in a high-level language with data-oblivious execution on a secure enclave. 
The DOT is designed to be built using only parameters related to the computation that are non-sensitive (such as data size). 
Because DOTs in DOVE are generated automatically, the client programmer does not need to learn the DOT language to write data-oblivious code. 
Once generated, the DOT is sent to the backend, where it is used to ``replay'' the same operations on the actual data (Section~\ref{sec:design:backend}).

What to include in the DOT semantics strongly influences the TCB size in the backend and DOVE's overall performance.
We designed the DOT semantics to follow the program counter (PC) model~\cite{pc_model}, at the granularity of primitive operations supported by the DOT.
That is, the structure of the DOT is similar to straight-line code where 
every operation is evaluated in the order it appears.
Conditionals, data-dependent loops, etc. must be emulated with predicated, bounded execution as described below.
We chose this design because while it can be difficult to transform normal programs to the PC model, it is generally much simpler to turn PC model programs into constant-time/data-oblivious programs.\footnote{We note that our current implementation implements data memory-trace obliviousness~\cite{MTO} in a simplistic fashion.  For example, if a data memory access has a sensitive address, we implement that access as a naive ``scan memory'' Oblivious RAM-style lookup.  Depending on parameters, future work can improve this using a poly-log overhead contant-time Oblivious RAM client~\cite{Raccoon,ZeroTrace}.}
For example, to convert an if-else style conditional to the PC model, a compiler (or similar) needs to convert the conditional to a predicated execution abstraction, which can be complex depending on whether the conditional is nested, etc.
However, converting predicated code into data-oblivious code usually entails simple transformations such as replacing point instructions with other side channel-resistant instructions such as \mintinline{nasm}{cmov}.
Thus, since the frontend is not in the TCB and the backend is in the TCB, we have pushed the complex program transformation tasks into the frontend, and therefore out of the TCB. 

Then, what primitive operations to include in the DOT semantics becomes a security/performance trade-off, because the cost to parse each operation in the DOT incurs non-negligible overhead in our current implementation (Section~\ref{sec:bend_impl}).
For example, DOVE might implement a transcendental function such as \mintinline{s}{sin} as a single primitive operation in the DOT or as a sequence of simpler operations in the DOT (such as bitwise operations).
The former design is higher performance but requires a larger TCB: the backend parses a single DOT operation and evaluates that operation using a dedicated data-oblivious implementation of \mintinline{s}{sin} in the target Instruction Set Architecture (ISA), e.g., x86-64. 
The latter has the opposite characteristics: the backend parses each bitwise operation yet only needs dedicated support to implement data-oblivious bitwise operations.
In these situations, we decide what operations to include in the DOT semantics on a case-by-case basis, described below and in Section~\ref{sec:design:frontend}.

We now discuss DOT semantics in more detail, using Figure~\ref{fig:diag_arch:fend} as a running example.
We break the discussion into two parts, first describing data creation and operations on said data, 
and second describing (data-oblivious) control flow. A formal EBNF grammar for the DOT can be found in the paper's extended version~\cite{dove_full}. 

\para{Data creation, types and operations.} 
 We first discuss variable declarations, types and primitive operations.

\subpara{Data types.}
When the frontend transcribes a program into a DOT, the DOT grammar only allows program inputs to be (1) fixed, concrete values or (2) pseudonyms. 
For example, in Figure~\ref{fig:diag_arch:fend}, the input \mintinline{s}{nrow(x)} of an R program (right) is translated into a concrete value \mintinline{haskell}{10} by the frontend, used as a fixed-loop bound in \textcircled{\raisebox{-0.9pt}{3}} for a DOT (left). This is possible because \mintinline{s}{nrow(x)} is fixed as \mintinline{s}{10} in line 1 of the original R program.
Likewise, a concrete value \mintinline{s}{0} that is being assigned to \mintinline{s}{x[i,1]} is transcribed with a prefix \mintinline{haskell}{#} in \textcircled{\raisebox{-0.9pt}{5}} in order to indicate that it is a concrete value.

The two basic types of pseudonyms are matrices and scalars, with matrices being composed of $m \times n$ scalar (i.e., numeric) elements. 
\textcircled{\raisebox{-0.9pt}{1}} in Figure~\ref{fig:diag_arch:fend} shows the definition of such a matrix, with $m = 10, n = 7$. 
Matrices are indicated with \mintinline{haskell}{$}, scalars with \mintinline{haskell}{

\subpara{Operations on data.}
Core functions comprise the set of primitive operations available to the DOT, including mathematical and logical operators (e.g. \mintinline{haskell}{+}, \mintinline{haskell}{==}), common mathematical functions (e.g. \mintinline{s}{exp}, \mintinline{s}{sin}), and summary operations (e.g. \mintinline{s}{sum}, \mintinline{s}{prod}).

There are two flavors of operations supported in the DOT, shown in first two rows of  Figure~\ref{fig:implemented_function}.
The Safe DOT/Core category contains operations deemed safe to operate on sensitive data in the backend.
Every operation in this set must be implemented data-obliviously by a compliant backend, i.e., its evaluation must result in operand-independent resource usage on the target microarchitecture (see Sections~\ref{sec:threat_model} and \ref{sec:sec_analysis}).
Each operation in this set has the following type signature: \emph{if at least one operand is a pseudonym, the result is a pseudonym.}
This is similar to taint algebras in information flow~\cite{DIFTSrini,lang_ift} where if one operand is tainted, the result is tainted.

The Unsafe DOT/Core category contains operations which the DOT deems not safe to operate on sensitive data.
For example, the \mintinline{c}{forloop} construct.
These operations are only allowed to take non-pseudonyms as operands.

Importantly, the selection which operations are marked Unsafe is a design choice.
An alternate set of DOVE semantics can specify a Safe variant of any Unsafe operation, subject to the constraint that the backend must support a data-oblivious implementation of said Safe operation.
Certain constructs, such as \mintinline{c}{forloop}, are difficult (and sometimes impossible) to implement data-obliviously with respect to their arguments---which motivates why we place them in the Unsafe category.
Thus, the trade-offs in deciding whether each operation is Safe vs. Unsafe are analogous to those for deciding which instructions should be made Safe vs. Unsafe in the Data-Oblivious ISA~\cite{oisa}.

\noindent To summarize, we have:
\begin{itemize}
\item[$\bullet$ \textbf{\emph{Rule 1:}}] If an operation's operand(s) are pseudonyms, the result is a pseudonym.
\item[$\bullet$ \textbf{\emph{Rule 2:}}] Safe operations may take pseudonyms or non-pseudonyms as inputs.  Safe operations must be implemented data obliviously by the DOVE backend.
\item[$\bullet$ \textbf{\emph{Rule 3:}}] Unsafe operations may only take non-pseudonyms as inputs.
\end{itemize}

This is analogous to the Data-Oblivious ISA policy \emph{Confidential data}$\nrightarrow$\emph{Unsafe instruction}, which is analogous to the classic policy \emph{High}$\nrightarrow$\emph{Low} in information flow. 
If a DOT follows the above rules, we call it a \emph{valid DOT}.
Whether a DOT is valid is checked before the DOT is evaluated by the backend (Sections~\ref{sec:design:backend}, \ref{sec:bend_impl}), and invalid DOTs are disallowed.

\para{Control flow.}
For reasons discussed above, the DOT disallows traditional control-flow constructs such as \mintinline{c}{if}, \mintinline{c}{while}, and \mintinline{c}{goto}, but supports predicated execution and bounded-iteration loops (similar to the program counter model~\cite{pc_model}).

\subpara{Bounded iteration.} The DOT provides a \mintinline{haskell}{forloop} iteration primitive that only allows non-sensitive/non-pseudonym predicates. 
This primitive further does not support infinite loops. \textcircled{\raisebox{-0.9pt}{2}} in Figure \ref{fig:diag_arch:fend} corresponds to the body of the loop.  In the example, \textcircled{\raisebox{-0.9pt}{3}} defines the bounds of the loop (from 1 to 10 in steps of 1), along with the loop index, \mintinline{haskell}{\1}. 
Loop indices are declared as non-pseudonyms.

We note that supporting \mintinline{haskell}{forloop} is purely a performance/DOT size optimization.
Equivalently, the loop could have been unrolled and the \mintinline{haskell}{forloop} construct removed.

\subpara{Predicated conditionals.} 
The DOT supports a \mintinline{haskell}{select} primitive that takes a pseudonym-typed predicate and returns one of two pseudonym operands based on the value of the predicate. \mintinline{haskell}{select} supports both scalar (i.e., logical \mintinline{s}{0} and \mintinline{s}{1}) and matrix predicates. Matrix predicates are transformed into element-wise select operations between the predicate and result/operand matrices.
Thus, the predicate and its operands must have the same dimensions. 

This flow is shown in \textcircled{\raisebox{-0.4pt}{4}} in Figure \ref{fig:diag_arch:fend}. 
This predicated execution model, similar to that used in prior work~\cite{Raccoon,practical_doprogramming}, facilitates data-oblivious branching.  
That is, once conditionals are re-written to \mintinline{s}{select}, it is relatively straightforward to further convert them to backend-specific data-oblivious operators such as \mintinline{nasm}{cmov} (Section~\ref{sec:background:data_oblivious_programming}).

Going back to our example, in \textcircled{\raisebox{-0.4pt}{4}}, the condition is a \mintinline{haskell}{<} comparison between \mintinline{haskell}{sensitive} at row \mintinline{haskell}{\1} (loop index), column 1, with the scalar value 0. This condition is a pseudonym, and thus it cannot be evaluated directly using the DOT alone. The \mintinline{haskell}{select} (\textcircled{\raisebox{-0.4pt}{5}}) uses this condition, and if it is true, returns a scalar 0.
Otherwise, it returns the value already in that location (\mintinline{haskell}{$1@(\1,1)}). 
This result value, stored in \mintinline{haskell}{

\subsection{Frontend}
\label{sec:design:frontend}

The frontend takes R program with non-sensitive parameters as input and outputs a DOT.
We develop our prototype frontend for R, but stress that the structure of the DOT is language-agnostic.
As in a traditional compiler stack, one could design a different frontend for a different language that likewise compiles into the DOT representation.

Before initialization, clients share non-sensitive information, such as names and dimensions of datasets, with each other.
The data within each dataset is considered sensitive and is not shared.
To create a DOT, a client sources the DOVE frontend, which loads the names and dimensions for each sensitive input and creates a pseudonym for each in the R environment. 
The client then runs their program, performing operations as normal.
Instrumentation in the R interpreter (see below) records each operation into the DOT, 
translating each dataset to primitives supported by the DOT semantics (e.g., scalar and matrix types).
Clients can access elements, assign new values, apply operators, and run functions, all while dealing only with pseudonyms. 
Because the frontend does not have the actual data, this transcription is sensitive data-oblivious by design. 

\begin{figure*}
    \footnotesize
    \centering
    \caption{DOVE functions/operations. Functions in group ``DOT/Core'' are implemented directly in the DOVE backend and are included in the DOT semantics. 
    Functions in the group ``Supplemental'' are implemented using operations in ``DOT/Core'' and exposed to the user as library functions. 
    Safe functions require a data-oblivious implementation in the backend as they may receive pseudonyms as operands.
    Unsafe functions do not require a data-oblivious implementation, but can only take non-pseudonyms (non-sensitive) data as operands.
    }
    \begin{tabular}{|lllllllll|}
    \hline
    Group & \vline & \multicolumn{7}{c}{Functions} \vline\\
   \hline
   Safe DOT/Core & \vline & \mintinline{s}{abs} & \mintinline{s}{sqrt} & \mintinline{s}{floor} & \mintinline{s}{ceiling} & \mintinline{s}{exp} & \mintinline{s}{log} & \mintinline{s}{cos}\\
   (in TCB) & \vline & \mintinline{s}{sin}  & \mintinline{s}{tan} & \mintinline{s}{sign} &\mintinline{s}{+} & \mintinline{s}{-}& \mintinline{s}{*} & \mintinline{s}{/}\\
   & \vline & \mintinline{s}{^} & \mintinline{s}{
   & \vline & \mintinline{s}{==} & \mintinline{s}{!=} & \mintinline{s}{|} & \mintinline{s}{&} & \mintinline{s}{!} & \mintinline{s}{all} & \mintinline{s}{any}\\
   & \vline & \mintinline{s}{sum} & \mintinline{s}{prod} & \mintinline{s}{min} & \mintinline{s}{max} & \mintinline{s}{range} & \mintinline{s}{is.na} & \mintinline{s}{is.nan}\\
   & \vline & \mintinline{s}{is.infinite} & \mintinline{s}{select} & \mintinline{s}{
   \hline
   
   Unsafe DOT/Core & \vline & \mintinline{s}{forloop} & \mintinline{s}{dim} & \mintinline{s}{[} & \mintinline{s}{[[} &  &  &\\
   (in TCB) & \vline & & & & & & &\\
   \hline
   Supplemental & \vline & \mintinline{s}{fisher.test} & \mintinline{s}{pchisq} &  \mintinline{s}{mean} &  \mintinline{s}{colMeans} & \mintinline{s}{colSums} & \mintinline{s}{rowMeans} & \mintinline{s}{rowSums}\\
   (not in TCB) & \vline & \mintinline{s}{is.finite} & \mintinline{s}{as.numeric} & \mintinline{s}{as.matrix} & \mintinline{s}{apply} & \mintinline{s}{lapply} & \mintinline{s}{unlist} & \mintinline{s}{which}\\
   & \vline & \mintinline{s}{data.frame} & \mintinline{s}{matrix} &  \mintinline{s}{split} &\mintinline{s}{pmin} & \mintinline{s}{pmax} & \mintinline{s}{nrow} & \mintinline{s}{ncol}\\
   & \vline & \mintinline{s}{len}  & \mintinline{s}{t} &&&&&\\
   \hline
   \end{tabular}
   \label{fig:implemented_function}
\end{figure*}

Our DOVE implementation ensures interface compatibility with base R in the implemented functions of the frontend. 
We use R's S3 method dispatch, as described in Section~\ref{sec:background:R} to overload functions in base R for pseudonyms.
This requires no modification to the R interpreter, as clients merely have to import the DOVE frontend in their existing programs; in most cases, no programmer intervention is necessary.

Figure~\ref{fig:implemented_function} lists all functions available to programmers. 
The Safe and Unsafe ``DOT/Core'' group of functions are those included in the DOT semantics (see previous section).
To provide a richer library for clients, we also provide a ``Supplemental'' group of functions which are built using only the operations in ``DOT/Core''. 
For example, \mintinline{s}{colSums} calls the DOT function \mintinline{s}{sum} in a loop over the columns of a matrix. 
We provide these functions to enhance the user programming experience and to 
show that our DOT functions are sufficient primitives to develop more complex functions. 
Note that the ``Supplemental'' functions do not add to size of the TCB. They do not require changes to DOT semantics and therefore do not change the backend implementation. 

\para{Construct-specific handling.}
We now describe how the frontend translates different R programming constructs to the DOT semantics from Section~\ref{sec:design:dot}. 

\subpara{Bounded iteration.} Native R's \mintinline{s}{for} loop is not DOT-aware, so it just repeats the body of the loop $m$ times. The frontend will naively record repeated invocations of the loop body every iteration; this results in the DOT size being proportional to complexity of the loop. Instead, the frontend automatically transforms such bounded loops to use the \mintinline{haskell}{forloop} DOT construct. 
As discussed in Section~\ref{sec:design:dot}, explicitly defining the \mintinline{haskell}{forloop} is purely a performance enhancement. 
In our testing, we observed a $>99 \%$ decrease in frontend runtime using the DOT's \mintinline{haskell}{forloop} loops over normal \mintinline{s}{for} loops for compute-heavy \(O(m^2)\)-complexity programs. 

We also note that many loops are written with early termination (e.g., \mintinline{s}{break}) conditions.
When a \mintinline{s}{break} statement is encountered, the frontend first examines a predicate associated with the break condition and its associated operations. 
Then the frontend performs a transformation 
to each statement in the loop to mask out architectural state updates, using \mintinline{s}{select}, once the break condition has been tripped.
This transformation is similar to those of prior works~\cite{FactLanguage,Oblivm}.

\subpara{Predicated conditionals.}
The frontend must translate conventional if-then-else structures into the predicated execution model supported by the DOT (Section~\ref{sec:design:dot}). 
For this, we implement an if-conversion transformation that is similar to prior works~\cite{Raccoon,practical_doprogramming}: an if-else with a sensitive predicate is converted into straight-line code where both sides of the if-else are unconditionally evaluated and a DOT \mintinline{haskell}{select} operator is used to choose the correct results at the end.

Our frontend automatically converts R \mintinline{s}{if} statements to use the \mintinline{haskell}{select} primitive (discussed in Section \ref{sec:design:dot}) in the DOT. The branch in Figure~\ref{fig:diag_arch:fend} is converted to \textcircled{\raisebox{-0.9pt}{4}} through this process. Updating the matrix value at the current position with either 0 or itself retains the semantics of the original \mintinline{s}{if} statement while making the operation explicitly data oblivious. 
The whole expression is then recorded into the DOT directly; since the frontend does not have access to the actual data, the DOT must necessarily record both sides of the condition. 

\subpara{Disallowed constructs.}\label{sec:design:frontend:disallowed}
Overall, the frontend's job is to translate R semantics into DOT semantics. 
Sometimes this is not possible, in which case the frontend signals an error.
We explain two such cases (which are also common issues in related work). 
First, the frontend does not allow loops where the predicate depends on a pseudonym.  
Second, the frontend does not allow running operations with unimplemented types 
e.g., string-based computation or symbol-based computation. 
For example, one genomic evaluation program named \mintinline{s}{geno_to_allelecnt} in Section~\ref{sec: perf_eval: eval_scripts} receives a matrix of characters as a sensitive input. 
This program calls string operations like substring search or string concatenation.

Importantly, mentioned before, the frontend may contain a bug that results in an invalid DOT that contains an illegal construct such as those mentioned above.
Such non-compliant DOTs are checked at parse time in the backend and rejected before being run (Section~\ref{sec:bend_impl}).

\para{Support for integrity protection.}
While our primary focus is privacy, DOVE can achieve integrity through SGX's attestation support.
Specifically, each client can run the frontend and generate the DOT locally.
These DOTs, or their hashes, can be checked against the DOT evaluated on the server side, by a DOVE backend that is attested to perform such checks.

\subsection{Backend}
\label{sec:design:backend}

The backend is a trusted SGX enclave that runs the DOVE virtual machine 
that parses the DOT and runs the instructions contained within on the clients' sensitive data. 
Code in the backend ensures that only valid DOTs are run (Section~\ref{sec:design:dot}), and includes
implementations of all operations in the DOT semantics, i.e, those listed under Safe and Unsafe ``DOT/Core'' in Figure~\ref{fig:implemented_function}.
Each client securely uploads (e.g., over TLS) the DOT of their R program. 
All clients additionally upload their shares of the sensitive dataset to the backend as well, in preparation for processing, as shown in Figure~\ref{fig:diag_arch:overview}.

The scope of DOVE is to block all non-speculative $\mu Arch$ side channels (Section~\ref{sec:threat_model}). 
For this purpose, the backend provides a data-oblivious implementation for operations in 
Safe ``DOT/Core'' of Figure \ref{fig:implemented_function}. 
To implement these operations, 
we rely on a subset of the x86-64 ISA and well-established coding practices~\cite{practical_doprogramming} for implementing constant-time/data-oblivious functions (see Section~\ref{sec:sec_analysis} for details). 
For example, we implement the \mintinline{s}{select} operation using the x86-64 \mintinline{nasm}{cmov} instruction, and all floating-point arithmetic functions are implemented using libfixedtimefixedpoint (libFTFP), a constant-time fixed-point arithmetic library created as a work-around for timing issues on floating-point hardware~\cite{FPU_leaky}. 

Importantly, what hardware operations (e.g., machine instructions) open \(\mu Arch\) side channels depends on the \(\mu Arch\).
For example, two x86-64 processors can implement \mintinline{nasm}{cmov} differently: one in a safe way, one in an unsafe way (e.g., by microcoding the \mintinline{nasm}{cmov} into a branch plus a move~\cite{oisa}).
DOVE is robust to new leakages found in specific \(\mu Arch\) because to block a newly discovered leakage, it is sufficient to make a backend change.
For example, if a vulnerability is found in \mintinline{nasm}{cmov}, the backend can opt to implement the DOT \mintinline{s}{select} operation using a CSWAP (bitwise operations) or other constructs.
\section{Implementation}
\label{sec:bend_impl}

\begin{figure}
    \centering
    \includegraphics[width=\columnwidth]{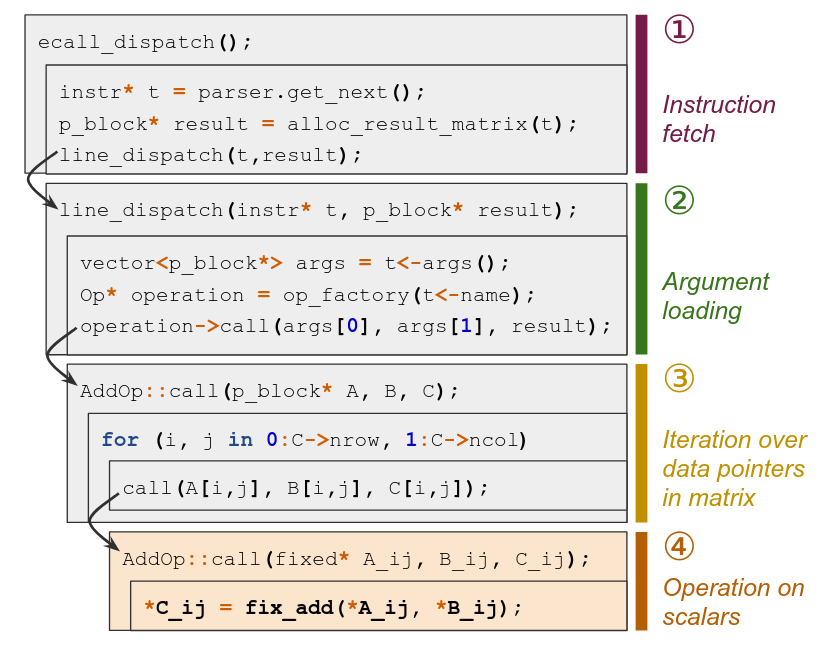}
\caption{A simplified graph representing the flow of \mintinline{c}{ecall_dispatch()} for the addition instruction \mintinline{s}{+}. The final, bold-face portion is the only block that dereferences sensitive data.}
\label{fig:add_pscode}
\end{figure}

We now discuss the backend implementation, the C++ codebase that evaluates DOT operations and forms the DOVE TCB.
We will rely on this description during our security analysis (Section~\ref{sec:sec_analysis}).
We eschew detailing our frontend implementation as it is outside the TCB. 
As noted previously, the backend code runs in an SGX enclave. The three functions that run in this secure enclave (or \mintinline{asm}{ECALL}s in SGX parlance) are associated with the three phases of the backend: loading sensitive data into secure memory (\mintinline{c}{ecall_load_data}), parsing the DOT into an abstract syntax tree (\mintinline{c}{ecall_parse}), and evaluating DOT operations based on said tree (\mintinline{c}{ecall_dispatch}). 

\para{Loading the data.} 
The enclave loads client data, as it is received, into the SGX EPC (Section~\ref{sec:background:sgx}).
The binary blobs received consist of 8-byte, little-endian \mintinline{c}{double} values along with metadata about the dataset format (\emph{e.g.}, tensor shape). The dataset is then copied into enclave memory via \mintinline{c}{memcpy}, and converted into a fixed-point integer representation. 
The dataset in memory is stored in a \mintinline{c}{p_block} data structure, which consists of a matrix of pointers to the scalar data values in memory and the matrix dimensions it represents.

\para{Parsing the DOT.} 
This phase involves recursive-descent parsing of the DOT into an abstract syntax tree (AST). Conceptually, this tree is akin to a list of instructions (see Figure~\ref{fig:diag_arch:fend}) to be interpreted by the enclave. 
Recall, the DOT is created by the frontend without access to sensitive data.
The AST, by extension, is not a function of sensitive data. 

Importantly, the parsing process verifies that the DOT complies with DOT semantics. 
Specifically, that no disallowed construct appears in the DOT (Section~\ref{sec:design:frontend}) and that each DOT operation type checks (Section~\ref{sec:design:dot}).
The latter ensures that pseudonyms cannot be downgraded to non-pseudonyms (Rule 1, Section~\ref{sec:design:dot}) and that pseudonyms are not passed as operands to Unsafe operations (Rule 3, Section~\ref{sec:design:dot}).

\para{Evaluating the DOT.} 
After loading datasets and parsing the DOT, DOVE is ready to run instructions from the DOT on the data. Broadly, this phase occurs in four steps per instruction across four functions in the backend. Figure~\ref{fig:add_pscode} depicts a simplified call graph for the addition instruction (\mintinline{s}{+}). Running different instructions entails a similar call graph.

During instruction fetch (\textcircled{\raisebox{-0.4pt}{1}} in Figure~\ref{fig:add_pscode}), the \mintinline{c}{ecall_dispatch()} function is the top-level call that fetches the next instruction from the DOT to be run and also allocates a placeholder datatype for use with the results of the instruction: in our running example, this is a matrix \mintinline{s}{C} of type \mintinline{c}{p_block*}. 
The argument loading step (\textcircled{\raisebox{-0.4pt}{2}}) loads pointers to instruction datatypes that form the arguments to the instruction. Our example loads two matrices: \mintinline{s}{p_block* A} and \mintinline{s}{p_block* B}. The iteration step (\textcircled{\raisebox{-0.4pt}{3}}) utilizes polymorphism to dispatch the backend operation corresponding to the instruction.

When one or more operands are matrices, as in Figure~\ref{fig:add_pscode}, we must perform the addition operation over all matrix elements. So, the final step is to iterate over all elements in the matrices and pass each element's pointer to a subcall to actually operate on the scalars in the matrix.

In this final step, control reaches a \emph{leaf function}, the highlighted, bottom level of the graph in Figure \ref{fig:add_pscode}.
This is the only step when the scalar elements of the sensitive data matrix are dereferenced and utilized in the operation. At this point in our example (\textcircled{\raisebox{-0.4pt}{4}}), we use the external libFTFP library to perform a data-oblivious operation, addition in our example, on actual scalar values. 
\section{Security Evaluation}
\label{sec:sec_analysis}

We first present our formal definition of security under the SGX adversary first discussed in Section~\ref{sec:threat_model}. 
We then argue that this security definition holds given our DOVE backend implementation, for the setup and instruction execution phases.

\subsection{Security Definition}
\label{sec:sec_analysis:sec_def}

In order to analyze the security properties of DOVE, we first formalize our security definition. We denote an execution of the R interpreter as \(R(S,D)\), where \(S\) is an R program, and \(D\) is the data on which the program \(S\) is run. The SGX adversary's view of \(R(S,D)\) (i.e., the leakage trace) is denoted \(\mu Arch\). As discussed in Section \ref{sec:threat_model}, this view includes hardware resource usage
at a fine spatial- and temporal-granularity. 
For example, the attacker can monitor contention for the cache or other hardware resources, the program runtime, etc. However, due to the virtual isolation provided by the SGX TEE, the view does not include the enclave memory itself.

From our analyses in Section \ref{sec: attack_examples}, it is clear that \( \mu Arch[R(S, D)] \neq \mu Arch[R(S, D')] \) for certain $D$ and $D'$. That is, the adversary's view of the \(\mu Arch\) side channels of the execution of the R interpreter on a program \(S\) is different for different datasets \(D, D'\). In practice, this means that the adversary can glean information from the datasets via these side channels, implying the computation for the given view is data dependent.

Now, we consider the notation for DOVE. We define the frontend as \(A: S \mapsto T_S\), a compiler \(A\) that translates \(S\) into a DOT \(T_S\). This computation is vacuously data-oblivious, since no data is passed as a parameter to \(A\). We then define the backend as \(V(T_S,D)\), a virtual machine that runs the DOT on the data. We aim to show that DOVE is secure against an SGX adversary, with the following definition.

\begin{defn}
We say a Data-Oblivious Virtual Machine backend \(V\) is \textit{secure in the SGX adversary model} if for any pair of datasets \(D\), \(D'\), and DOT \(T_S\) compiled from a program \(S\), we have the following equation.
\[
    \mu Arch[V(T_S, D)] = \mu Arch[V(T_S, D')]
\]
where \(\mu Arch\) denotes the adversary's view in the SGX adversary model.
\end{defn}

This is equivalent to a non-interference property, where high (sensitive) state is the backend input data and low (non-sensitive) state is other architectural state in the processor (across all running programs).
We show that the DOVE implementation meets the above security definition through an analysis of the flow of sensitive data through the backend and compiled object code the backend uses.

\subsection{Security Argument}

Our evaluations were performed on a machine with an Intel Skylake Core i3-6100 CPU, 1 TB HDD, and 24 GB of RAM, of which 19.37 GB was allocated to the SGX enclave. The machine was running Ubuntu 18.04.4 LTS and SGX software version 2.9.1 with EPC paging support. Thus DOVE's memory is not limited to EPC size, but this mechanism adds performance overhead when it is required. 
We analyze this further in Section~\ref{sec: perf_eval}. 

The frontend ran under R interpreter version 3.4.4, and the backend was compiled against \texttt{g++}, toolchain version \texttt{7.5.0-3ubuntu1\textasciitilde18.04}. 
For our security analysis, we ran DOVE without SGX enabled for easier inspection of potential side channels. Our security evaluation related to side channels is independent of SGX, with the enclave technology being an implementation choice to guard against direct introspection/tampering by supervisor software. 
We provide a performance evaluation of DOVE in Section~\ref{sec: perf_eval}.

The backend forms our trusted computing base, and it consists of 7,001 lines of code, of which 4,295 lines is the libFTFP library which we adopt from prior work~\cite{FPU_leaky}.
Since the frontend has no access to sensitive data, the security evaluation of DOVE reduces to the evaluation on the remaining 2,706 lines of code in the backend. 
We now examine each step of DOVE's workflow prior to the leaf function call (described in Section~\ref{sec:bend_impl}).
Finally, Section~\ref{sec:sec_analysis:leaf_functions} examines the leaf functions.

\para{Creating the DOT.} 
Each client runs their R program on their local frontend to produce a DOT. The adversary can only learn non-sensitive information (e.g., dataset dimensions) explicitly given to the frontend.

\para{Transferring the data and DOT.}  
After DOT creation, the DOT and sensitive data is sent by the user to the server through a secure channel whose endpoint is within the TEE~\cite{aublin2017talos}.  The secure channel and SGX-provided attestation ensures that the correct DOT is run and that the data is privacy/integrity-protected in transit.  
In the case of multiple users submitting data and DOTs, this process is applied to each user, after which the DOTs are hashed and compared to ensure the computation is consistent with that requested by each user (also see Section~\ref{sec:design:frontend}).

\para{Loading the data.} 
Once datasets arrive, the SGX backend stores the data in the enclave by passing the datasets through \mintinline{c}{ecall_load_data}. As mentioned in Section~\ref{sec:bend_impl}, each dataset is a binary blob which is copied into enclave memory via \mintinline{c}{memcpy}.  
This operation consists of a sequence of data-oblivious \mintinline{asm}{mov} operations. 
Prior to the copy, we convert floating point values in each sensitive dataset to fixed point numbers (to be compatible with libFTFP~\cite{FPU_leaky}). 
This conversion process is also data-oblivious; further, the number of bits in the fixed-point representation is independent of the underlying value. 

\para{Parsing the DOT.} 
As noted in Section~\ref{sec:design:dot}, 
the DOT contains no sensitive information. 
By extension, the DOT parsing phase---whereby the DOT is parsed into an AST---cannot leak sensitive information.
Recall that the parsing process also ensures the DOT complies with DOT semantics (Section~\ref{sec:bend_impl}), which ensures that Rules 1 and 3 are enforced (Section~\ref{sec:design:dot}).
The DOT grammar is simple, so type-checking the DOT is also simple.

\para{Evaluating the DOT.} 
This phase occurs in four steps, as shown in Figure \ref{fig:add_pscode}: instruction fetch, argument loading, matrix iteration and operation on dereferenced sensitive data. 
The first three phases do not perform operations on the underlying sensitive data and only operate based on the DOT, which is a function of non-sensitive information as discussed above. 
Specifically, until the runtime reaches the operation in the leaf function, pointers to the sensitive data are passed around, but the sensitive data values are never accessed. 

\subsection{Leaf Functions}
\label{sec:sec_analysis:leaf_functions}

Based on the above discussion, only leaf functions read and modify sensitive data.
Thus, we now scrutinize whether these leaf functions enable our security guarantee, i.e., uphold Rule 2 from Section~\ref{sec:design:dot}.
For this, we manually disassemble and analyze every binary object file associated with DOVE functions, and verify that the subset of instructions which operate on sensitive data are instructions that do not create \(\mu Arch\) side channels as a function of their operands. 
The following analysis applies well-established principles for writing constant-time and data-oblivious programs (Section~\ref{sec:background:data_oblivious_programming}).

We first analyze the leaf function instructions that take sensitive data as operands.
These instructions are shown in Figure~\ref{fig:opcode_table}.
We determined this set by inspecting instruction dependencies in the \mintinline{sh}{objdump} disassembly.
All but one of the opcodes in Figure~\ref{fig:opcode_table} is considered to be a data-oblivious instruction by libFTFP, our constant-time fixed-point arithmetic library. 
We refer to its authors' analysis for its security~\cite{FPU_leaky}. 
The one instruction not found in libFTFP, \mintinline{asm}{cmovne},
is used for conditional moves of sensitive data in the backend. 
This instruction is likewise shown to be data oblivious in \cite{Raccoon}. 
We further verify that the above instructions use the direct register addressing memory mode for each operand, if the value stored in the register for that operand is sensitive (which also follows standard practice for writing data-oblivious code).\footnote{x86-64 operands can utilize one of several flavors. For example, \texttt{rax} denotes a register file read and \texttt{[rax]} denotes a memory de-reference. The former is considered safe for use in constant-time/data-oblivious programming, while the latter creates memory-based side channels.}
Thus, we conclude that the machine instructions operating on sensitive data in the backend do not create \(\mu Arch\) side channels.

Beyond the instructions in Figure~\ref{fig:opcode_table}, there are other instructions in the leaf functions that \emph{do not} operate on sensitive data.
Examples include jumps to implement loops with non-sensitive iteration counts, checks to validate dimensions on operations, sanity checks for \mintinline{c++}{nullptr}, and instructions associated with implementing polymorphism. Some of these are not data oblivious (e.g., jumps), but do not impact security because they operate on non-sensitive data such as matrix dimensions.

To further corroborate our static security analysis, we also looked at runtime instruction statistics. 
We used the branch-trace-store execution trace recording~\cite{intel2018intel} of the DOVE backend execution, varying the input data.
We found that the sequence of non-speculative dynamic instructions executed was independent of the data passed to the backend:
that is, the backend satisfies the PC model~\cite{pc_model}.
Security follows from these two analyses: (a) that the backend follows the PC model and (b) that each individual instruction that operates on sensitive data consumes operand-independent hardware resource usage (previous paragraphs).
Additional details regarding dynamic security analysis of the DOVE backend can be found in the paper's extended version~\cite{dove_full}.

\begin{figure}
    \centering
    \caption{All x86-64 opcodes that operate on sensitive data in the leaf functions of DOVE. Those marked with * are those not found in libFTFP.}
    \resizebox{0.9\columnwidth}{!}{%
    \begin{tabular}{|llllll|}
    \hline
    \texttt{add} & \texttt{and} & \texttt{cdqe} & \texttt{cmovne}* & \texttt{cmp} & \texttt{imul} \\
    \texttt{lea} & \texttt{mov} & \texttt{movabs} & \texttt{movsd} & \texttt{movsx} & \texttt{movsxd} \\
    \texttt{movzx} & \texttt{mul} & \texttt{neg} & \texttt{not} & \texttt{or} & \texttt{pop} \\
    \texttt{push} & \texttt{sar} & \texttt{sbb} & \texttt{seta} & \texttt{setae} & \texttt{setbe} \\
    \texttt{sete} & \texttt{setg} & \texttt{setl} & \texttt{setle} & \texttt{setne} & \texttt{shl} \\
    \texttt{shr} & \texttt{sub} & \texttt{test} & \texttt{xor} &  & \\
    \hline
    \end{tabular}
    }%
    \label{fig:opcode_table}
\end{figure}
\section{Experimental Evaluation}
\label{sec: perf_eval}
We now turn to the experimental evaluation of DOVE in three areas: (1) correctness, (2) expressiveness, and (3) computational efficiency.
It is necessary to provide some evidence that computed values are correct, at least for a basic collection of computations. Since DOVE works with a subset of R, it is also important to demonstrate that it can code enough interesting cases to be worthwhile. Moreover, DOVE computations must sufficiently limit computational overhead. 
We carry out the validation via two case studies, using the same machine configuration that we introduced in Section~\ref{sec:sec_analysis} for experimental evaluation. The first echoes prior work~\cite{SGXBigMatrixDOVE} by coding and analyzing applications of the PageRank algorithm~\cite{PageRank}.
The second examines a suite of programs~\cite{eva_chan} for genomic analysis and a case study using it for the analysis of honeybee genomes~\cite{bee_study}. It is easier to work with this type of data than, say, genomic data of people with bipolar disorder, while
it illustrates similar issues of scale and the potential value of controlled data sharing.

For correctness, we confirm that what we get from DOVE is the same as what we would get from R. That is, using the notation of Section \ref{sec:sec_analysis:sec_def}: $Q[V(T_S, D)] = Q[R(S, D)]$ where $Q$ denotes the calculated output for a given execution. 
For expressiveness, we demonstrate that we can conveniently create DOTs from R code for each case study. As such, we devote most of the section to the evaluation of computational efficiency.

One run of our performance benchmark is as follows. We first record the runtime of vanilla (insecure) R with data and a program. Then, we run the DOVE frontend on the same program, generating the DOT and writing it to disk. We then initialize the backend, read in the DOT, parse it, and execute the DOT instructions. Our evaluation of the DOVE implementation discusses two measures. First, we wish to consider if our frontend primitives are sufficient to express complex programs. Second, we examine the performance of DOVE when compared to its base R counterpart. 

To highlight the overheads inherent to SGX and libFTFP, the external data-oblivious fixed point library, we ran performance benchmarks on three configurations of DOVE: (1) backend outside an SGX enclave and without libFTFP, (2) backend outside an SGX enclave and with libFTFP, and (3) backend inside an SGX enclave and with libFTFP (our default configuration).
SGX-related overheads include SGX's memory encryption and access protections that isolate the enclave from the rest of the machine~\cite{sgx-explained}.
The libFTFP instructions' relative performance overhead is measured against its Streaming SIMD Extensions (SSE) counterpart; the overhead varies depending on the instruction, ranging from 1.2$\times$ for \mintinline{nasm}{neg} (operand negation) to 208$\times$ for \mintinline{nasm}{exp} (exponential function evaluation)~\cite{FPU_leaky}.

\subsection{PageRank} 
We begin with an introductory case study on the PageRank algorithm that is used as a case study on a custom data-oblivious programming language~\cite{SGXBigMatrixDOVE}. A large proportion of this algorithm is composed of matrix multiplications, which other works choose as primary performance benchmarks~\cite{Raccoon,Ghostrider}.

We use the Wikipedia vote network (WikiVote)~\cite{wikivote} and Astro-Physics collaboration network (ca-AstroPh)~\cite{astroarxiv} datasets from the Stanford Network Analysis Project~\cite{snapnets} for this case study. Both datasets are converted to adjacency matrices, where WikiVote has 7,115 nodes (\(\approx\) 405 MB) and ca-AstroPh has 18,772 nodes (\(\approx\) 2.8 GB).
Figure~\ref{fig:page_rank_table} shows the runtimes for the PageRank algorithm on two datasets where the DOVE frontend took on average of 3.34 seconds for each dataset.
Note that we evaluate several configurations for the DOVE backend (SGX, libFTFP) as discussed before. 

Vanilla R ran faster than DOVE, even without the SGX enclave and without libFTFP.
This is expected for several reasons. First, R uses the highly-optimized Fortran BLAS library for matrix multiplication, while DOVE does not. 
Second, DOVE code (with or without libFTFP) disables compiler vectorization for safety reasons.
Finally, DOVE uses the data-oblivious x86-64 \mintinline{nasm}{cmov} for any conditional statement on sensitive data whereas the R interpreter is written with unsafe branch statements (Section~\ref{sec: attack_examples}).

Enabling libFTFP increases runtime overhead of DOVE by around 6$\times$, and enabling SGX on top of libFTFP incurs 3$\times$ additional overhead. 
The PageRank implementation shows that DOVE is expressive enough to handle a common data-processing algorithm without severe performance degradation. 

\begin{figure}
    \caption{Runtimes for running PageRank algorithm on different configurations. All measurements are in seconds, and the measurements are sums of frontend and backend runtimes. The frontend took on average of 3.34 seconds.
    }
    \centering
    \footnotesize
    \begin{tabular}{l|l|l}
        Configurations & WikiVote & ca-AstroPh \\
        \hline
        Vanilla R & 6.91 & 46.88 \\
        DOVE w/o libFTFP, w/o SGX & 23.70 & 122.18 \\
        DOVE w/ libFTFP, w/o SGX & 137.00 & 951.62 \\
        DOVE w/ libFTFP, w/ SGX & 509.04 & 2,254.46 \\
    \end{tabular}
    \label{fig:page_rank_table}
\end{figure}

\subsection{Genomic Analysis}
\label{sec: perf_eval: eval_scripts}
To further validate DOVE, we work with an application that performs a controlled study on honeybee genomic data~\cite{bee_study}. 
The study relies on R code drawn from a set of 13 genetics research programs~\cite{eva_chan} that implement important statistical measurements found in the literature~\cite{ec_ref0,ec_ref1,ec_ref2,ec_ref3},
totaling 478 lines of R code~\cite{eva_chan}. These programs, in addition to the coding of PageRank, constitute a practical illustration of the expressiveness of DOVE. 
The paper's extended version~\cite{dove_full} provides more details about the programs' applications to genomics. 

Using DOVE, we were able to transform (in the frontend) and run (in the backend) 11 out of the 13 evaluation programs, totaling 326 lines of R code. 
The first program that we could not implement, \mintinline{s}{geno_to_allelecnt}, works on character data instead of numeric data, and as such is not supported by the current types available in the DOT. The second program, \mintinline{s}{gwas_lm}, performs a Genome-Wide Association Study (GWAS) using support in R for linear models. We were not readily able to implement this; R provides parameters to models as a formula of symbols, not values. DOVE currently does not support this paradigm, but we believe that DOVE can be extended to do so in the future.

Ten of the remaining 11 evaluation programs were automatically transformed by the frontend into data-oblivious code.
Only one program, \mintinline{s}{LD}, required manual intervention, as it was written entirely in a data-dependent style.
For this program we: (1) replaced some functions that are intrinsically data-dependent with data-oblivious primitives and (2) changed lines that required sensitive data-dependent array indexing with worst-case array scans.
Future implementations could alternatively use an oblivious memory, e.g., \cite{ZeroTrace}, to avoid such worst-case work.

We utilize the dataset from the honeybee study~\cite{bee_study} 
to perform performance benchmarking. We run the full 2,808,570 \(\times\) 60 (\(\approx\) 1.3 GB) dataset for all programs with space complexity of \(O(m * n)\) where \(m\) is the number of rows and \(n\) is the number of columns. However, some of the evaluation programs could not run on this dataset due to machine limitations. Specifically, some programs with space complexity of \(O(m^2)\) refuse to run even in vanilla R at full size. To address these limitations, we run a subset of programs with the first 10,000 rows of the honeybee dataset. Some related work also runs performance benchmarks on genomic data with similar sizes to that of our reduced dataset~\cite{chen2016princess,chen2017presage,sadat2017safety}.

To normalize benchmark results run on datasets of different sizes, we present a relative overhead metric: runtime for DOVE (DOT generation, disk reading/writing, DOT evaluation) divided by runtime in vanilla R. This relative overhead metric is shown as stacked bar graphs in Figure~\ref{fig:performance_graphs}.\footnote{Numbers for each program's absolute runtime are given in the paper's extended version~\cite{dove_full}.}
Each part of the bar represents the overhead contributed by a component of the backend, categorized by three factors: the DOVE runtime's data-oblivious implementation itself, constant-time fixed point operations (libFTFP), and the use of the SGX enclave. Overall, each factor provides additional security at the cost of increased overhead. We separate our programs into two bins: programs that run on the full honeybee dataset, and programs that run on a reduced dataset due to machine limitations (marked with * across the subfigures). 

The min/avg/max size overhead of each DOT relative to its R script is 0.284x/10.8x/105x. 
Note, the DOT may be smaller than the original program because of the DOT instruction set.
We expect that the DOT can be significantly compressed.
Case in point, the current DOT is represented in ASCII which is space inefficient.

We now provide more detailed analysis for several programs with noteworthy performance characteristics.

\para{Programs with quadratic space complexity.}
The relative overhead with DOVE is 120.7\(\times\) against vanilla R on average for programs \mintinline{s}{EHHS}, \mintinline{s}{iES}, and \mintinline{s}{LD}.
These three programs run statistics based on pairwise SNPs, i.e., a row is compared to each other row in the dataset. They operate in \(O(m^2)\) space, or, quadratic in the number of rows \(m\). The large relative overhead in the base DOVE implementation for \mintinline{s}{iES} and \mintinline{s}{EHHS} is due to data-oblivious transformations. 
Namely, the vanilla R versions of these programs benefit from early \mintinline{s}{break}s in the loop body that occur depending on sensitive values. DOVE does not directly allow such behavior for security reasons. Hence, the backend must iterate through the entire matrix, regardless of the data, causing potentially high overhead.

\para{Statistical programs.} 
The programs \texttt{hwe\_chisq} and \texttt{hwe\_fisher} each call a base R statistics function: \mintinline{s}{pchisq} (Chi-Square distribution) and \mintinline{s}{fisher.test} (Fisher's exact test), respectively. The program \texttt{snp\_stats} calls both functions. In base R, the implementation of \texttt{fisher.test} is written in R itself whereas \mintinline{s}{pchisq} is written in C. We implement both 
as supplemental group functions in R (Figure~\ref{fig:implemented_function}), to provide a fair comparison and to reduce TCB size.
When called, the frontend will convert the call into a series of equivalent DOT operations.

We note that, to achieve data obliviousness, our implementations of these functions are somewhat different than their vanilla R counterparts. 
For instance, computing a factorial of a sensitive value is intrinsically data dependent, but it is required to compute Fisher's exact test (in R, \mintinline{R}{fisher.test}). 
To implement factorial data obliviously, we implement it as an oblivious table lookup over a pre-determined domain of inputs, noting that other data-oblivious implementations are possible.

While \texttt{hwe\_chisq} has reasonable performance overhead given our data-oblivious implementation of \texttt{pchisq}, both \texttt{hwe\_fisher} and \texttt{snp\_stats} show large performance overheads. 
These programs call the \texttt{fisher.test} function \(O(m)\) times.
The insecure version of this function takes \(O(n)\) time.
Our data-oblivious implementation takes \(O(n^2)\) time 
due to inefficient oblivious-memory reads.
As mentioned before, a more efficient oblivious-memory primitive would reduce overhead.

\para{Remaining programs.}
The remaining programs do not incur a significant performance penalty, as both the insecure and data-oblivious codes run in $O(m)$ time. 
The average overhead with DOVE is 28.3\(\times\) relative to vanilla R for these programs. 
One program, \mintinline{s}{allele_sharing} (in Figure~\ref{fig:perf_graph_large}), has a notably larger performance overhead than others when running inside the SGX enclave.
We believe this is due to EPC paging costs.
Specifically,
this program has a larger working set size than SGX has EPC/PRM (2~GB vs. 64-128~MB). 
It further makes column-major traversals for a matrix that is stored in row-major order in memory, which leads to low spatial locality and therefore, we hypothesize, a high EPC fault rate.

\begin{figure}[t]
\centering
\begin{subfigure}[t]{\linewidth}
    \centering
    \includegraphics[width=\linewidth]{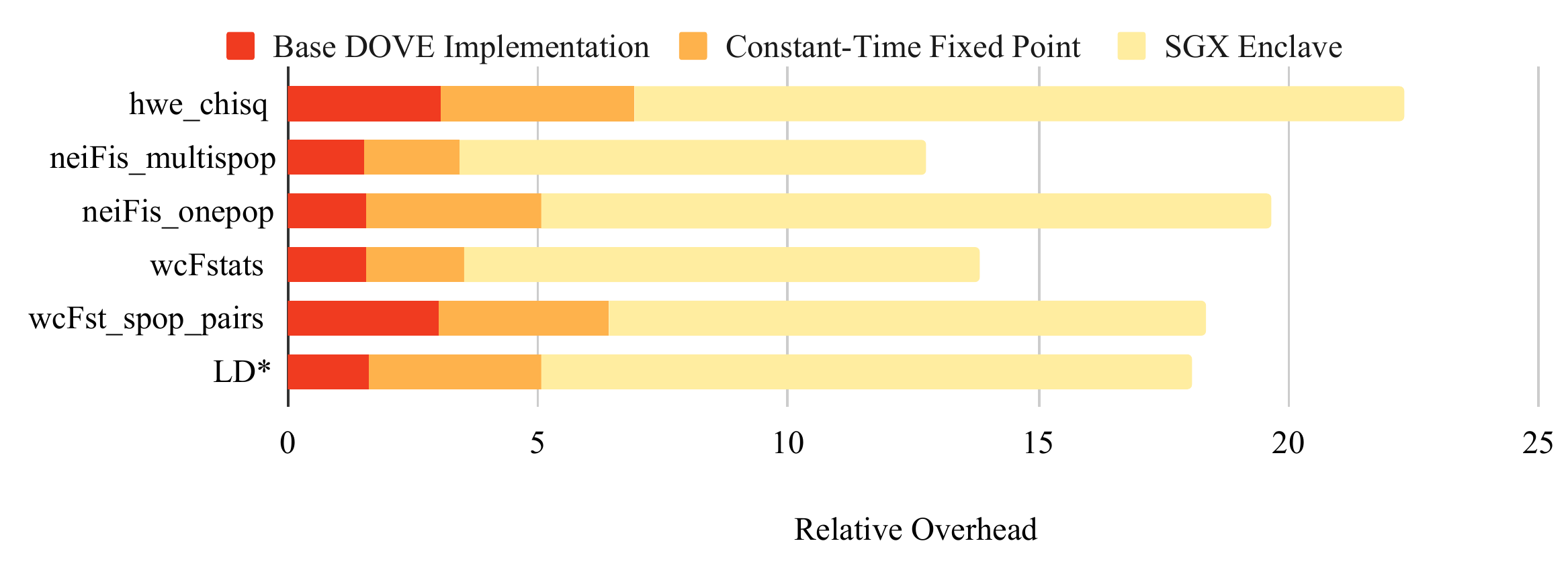}
    \subcaption{Programs with less than \(25\times\) relative overhead.}
    \label{fig:perf_graph_small}
\end{subfigure}
\begin{subfigure}[t]{\linewidth}
    \centering
    \includegraphics[width=\linewidth]{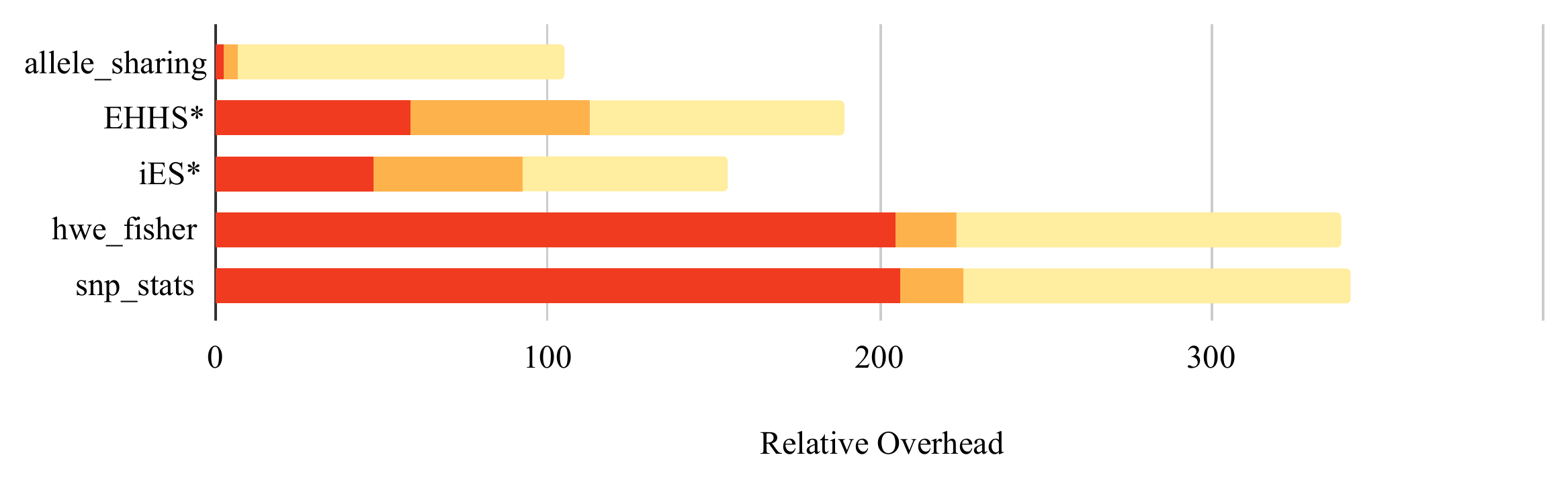}
    \subcaption{Programs with greater than \(25\times\) relative overhead.}
    \label{fig:perf_graph_large}
\end{subfigure}
    \centering
    \caption{Performance evaluation results for the evaluation programs. Each stacked bar represents a measurement for each program. Each stack represents relative overhead of DOVE against vanilla R caused by generic data-oblivious computation, libFTFP and SGX from left to right. Programs marked with * run on reduced dataset due to machine limitations.}
    \label{fig:performance_graphs}
\end{figure}
\section{Discussion and Future Work}
\label{sec: discussion}

\para{Current prototype limitations.}
The DOT has been designed to provide functionality for real-world data science tasks. 
Some features such as multi-threading and networking, that are present in general-purpose languages, are not currently supported in either the DOT or DOVE more generally. This is not fundamental. 
Additional functionality can be added to the DOT, as long as there exists a data-oblivious implementation of said functionality. 

\para{Handling loop bounds (and related constructs) that depend on sensitive data.}
In the current DOVE prototype, loop bounds (and related constructs such as recursion depth) must be a function of non-sensitive data.
For example, we consider matrix dimensions to be non-sensitive and matrix dimensions determine loop bounds in our evaluation scripts.
An interesting direction for future work would be to add either static or dynamic program analysis to enable such control-flow information to be a function of sensitive data.
For example, if a given loop iterates $i_1$ or $i_2$ times depending on a sensitive value, one would like an analysis to discover $i_1$ and $i_2$, set the loop's bound to $\mathsf{max}(i_1, i_2)$ and add the instrumentation from Section~\ref{sec:design:frontend} to mask out architectural state updates when the actual input requires fewer loop iterations.

\para{Possible performance optimizations.}
Finally, as our primary objective was to demonstrate a proof-of-concept of DOVE's security benefits, we believe
many performance optimizations are possible. 
For example, there are performance-optimized data-oblivious implementations of several key primitives (e.g., sensitive array lookup~\cite{ZeroTrace}, matrix multiply~\cite{SGXBigMatrixDOVE}) which could be integrated into our backend to improve performance without changing the DOVE architecture. 
Finally, as mentioned in Section~\ref{sec:design:dot}, users can add frequently-used routines as DOT primitives implemented in the backend, to trade-off performance and TCB size.

\section{Related Work}
\label{sec: related_work}

\subsection{SGX Programming}
Our work is related to prior efforts in running/partitioning/managing general purpose applications in SGX~\cite{haven,graphene_sgx,scone,panoply,glamdring,trustjs,scriptshield,ryoan,ZeroTrace,SGXBigMatrixDOVE,felsen2019secure}.
The four most relevant axes for comparison are: 
(i) whether the application running is untrusted, 
(ii) whether the proposal runs interpreted code such as R,
(iii) what is the threat model (in particular, does it include defense against \(\mu Arch\) side channels) and
(iv) whether the proposal requires a new custom programming language.
We show a comparison along these axes in Figure~\ref{tab:related_work}.
The takeaway is that no prior proposal, to our knowledge, simultaneously runs (i) untrusted code, (ii) high-level interpreted code such as R, (iii) provides broad protection against \(\mu Arch\) side-channel attacks, (iv) supports existing languages.
Moreover, our work makes a distinct conceptual- and design-level contribution, namely to orchestrate computation through existing high-level languages without showing those languages the sensitive data.

The works most similar to our proposal are TrustJS~\cite{trustjs} and SGXBigMatrix~\cite{SGXBigMatrixDOVE}. 
The former runs untrusted JavaScript code but assumes a weaker adversary that cannot monitor fine-grain application behavior over, e.g., side channels.
The latter provides a data-oblivious matrix API, but requires programmers to adopt a custom scripting language for performing computation.
We view this work as complementary: our work strives to enable general-purpose data oblivious computing on \emph{existing high-level languages}, but could benefit from the performance optimizations made to matrix computations in SGXBigMatrix.

\subsection{Data-Oblivious Programming}
There is a rich literature that studies how to write and run different applications in a data-oblivious fashion on today's ISAs.
For example, application-centric works propose data-oblivious cryptography~\cite{curve25519, poly1305}, machine learning~\cite{SGXBigMatrixDOVE,OblivML}, databases~\cite{Opaque,OblivSQL,Oblix}, memory and datastructures~\cite{ZeroTrace,obliviate}, general purpose code~\cite{pc_model,Raccoon,practical_doprogramming, felsen2019secure,sharemindHI}, utilities~\cite{ObliviousUtils} and floating point functions~\cite{FPU_leaky}.
Many of these~\cite{IronDOVE,ZeroTrace,Opaque,OblivSQL,Oblix,SGXBigMatrixDOVE,OblivML,Raccoon,felsen2019secure,sharemindHI} were designed for SGX-enabled applications to block the \(\mu Arch\) side channels discussed in Sections~\ref{sec:background:side_channels}-\ref{sec:background:sgx}.
Programming language, compiler and runtime works study how to write (e.g., \cite{Lobliv,FactLanguage}) and compile (e.g., \cite{Oblivm,ZahurCompiler,tiny_garble}) programs to software circuits.
ISA abstractions study how to design interfaces usable by both software designers and hardware architects to uphold data-oblivious security guarantees~\cite{oisa}.
These efforts are backed on the theory side by studies on how to run different algorithms and data structures in the circuit model~\cite{path_oram, OBFS, GraphSC, OStableMatching, CacheObliviousSort, Oblivm,tiny_garble,ZahurOakland,ODS}.

Our work differs from these application-centric works by targeting high-level interpreted programming stacks such as R as opposed to low-level C.
As discussed in Section~\ref{sec: attack_examples}, R introduces significant new challenges in establishing confidence that code is actually data-oblivious.
Our work differs from the PL, runtime, compiler work because it focuses on hardening mostly-unmodified R code, as opposed to creating a new end-to-end stack with a custom language; our work differs from ISA work, such as the Data-Oblivious ISA extensions~\cite{oisa} (OISA), by not requiring ISA-level changes to the underlying machine.
Finally, our work is complementary to data-oblivious algorithm and data structure design, as our backend can leverage these algorithms to implement specific operations.

\subsection{Privacy-Preserving Genomics}
Our case study is based on the genomics of honeybees collected from three different locations~\cite{bee_study}. Genomics has been a promising test case for privacy-preserving application of SGX before in other areas. These include at least the privacy-preserving computation of admixtures~\cite{chen2016premix}, Genome Wide Association Studies (GWAS)~\cite{sadat2017safety}, and analysis of rare diseases (viz. Kawasaki disease)~\cite{chen2016princess}.  There is also an SGX-based study of privacy-preserving queries on genomic data~\cite{chen2017presage}. There is a survey \cite{aziz2017privacy} that discusses approaches to genomic privacy based on SGX, on cryptography, and on a hybrid of both. 

\begin{figure}
\caption{\label{tab:related_work} Related works on application partitioning/management in SGX.
}
\footnotesize
\begin{tabular}{p{2.3cm}p{.8cm}p{.9cm}p{1.25cm}p{1.25cm}}
Name                                                & Untrusted Apps?               & Interpreted Code? & Blocks $\mu$arch Side Channels? & Supports existing languages? \\
\hline
Haven~\cite{haven}                                  & \xmark                        & \cmark            & \xmark                & \cmark              \\
Graphene-SGX~\cite{graphene_sgx}                            & \xmark                        & \cmark            & \xmark                & \cmark              \\
Scone~\cite{scone}                                  & \xmark                        & \cmark            & \xmark                & \cmark              \\ 
Panoply~\cite{panoply}                             & \xmark                        & \xmark            & \xmark                & \cmark              \\
Glamdring~\cite{glamdring}                          & \xmark                        & \xmark            & \xmark                & \cmark              \\
TrustJS~\cite{trustjs}                              & \cmark                        & \cmark            & \xmark                & \cmark              \\ 
ScriptShield~\cite{scriptshield}                    & \xmark                        & \cmark            & \xmark                & \cmark              \\
Ryoan~\cite{ryoan}                                  & \cmark                        & \xmark            & \xmark                & \cmark              \\
SGXBigMatrix~\cite{SGXBigMatrixDOVE}                    & \cmark                        & \cmark            & \cmark                & \xmark              \\
ZeroTrace~\cite{ZeroTrace}                          & \xmark                        & \xmark            & \cmark                & N/A                 \\
Felsen et al.~\cite{felsen2019secure}                          & \cmark                        & \xmark            & \cmark                & N/A                 \\
DOVE (This paper)                                   & \cmark                        & \cmark            & \cmark                & \cmark              \\
\end{tabular}
\end{figure} 

\section{Conclusion}
\label{sec: conclusion}
DOVE offers an approach to achieve data-oblivious computation within a TEE for programs originally written in languages with complex stacks such as R. The approach takes as input a high-level program and transforms it to an intermediate representation (DOT) that can be more easily reasoned about with respect to providing data obliviousness on a 
constrained TCB. This gives the advantage of being able to program in a familiar and convenient language while providing a very strong security guarantee. The trade-off for achieving these benefits, in general, is limits on the high-level programs that can be processed. We have demonstrated a design and implementation that can cover a significant range of programs with efficiency that is an acceptable trade-off for the benefits. This provides a foundation for future study using our methodology, such as expanding to richer high-level programming constructs and languages.

\section*{Acknowledgments}
This work was supported in part by NSF CNS 13-30491 (THaW), NSF CNS 19-55228 (SPLICE) and NSF CNS 18-16226 (OISA). The views expressed are those of the authors only.

\bibliographystyle{plain}
\bibliography{dove,chris,security,deep_learning}

\ifdefined\isconf 
\else 
\appendix
\subsection{The DOT Language}
\label{app:write_in_dove}

We provide a full grammar for the DOT language in extended Backus\textendash Naur Form (EBNF) in Figure~\ref{fig:dove_grammar}. Note that clients who write programs in high-level languages do not need to know the precise semantics of the full DOT language, as their frontend will handle this transcription transparently.

\begin{figure*}
    \caption{The grammar for the DOT language, presented in extended Backus–Naur form. \mintinline{ebnf}{digit} consists of 0-9, \mintinline{ebnf}{nonzero-digit} of 1-9, and \mintinline{ebnf}{alpha} of A-Z and of a-z.
    }
\footnotesize
\begin{minted}[breaklines,breaksymbolleft={},breakindent=12pt]{ebnf}
loop = 'forloop' index-var '\n' {instr} 'endloop' index-var ;

instr = (def-matrix | scalar-instr | edit-instr | select-instr) '\n' ;

def-matrix = create defn end ;
create = 'def' ' ' ['const' ' '] matrix ' ' length ' ' length '\n' ;
defn = rows | dataset | matrix-instr | bind-instr ;
rows = row {row} ;
row = '\t' 'row' ' ' natural ' ' scalar {' ' scalar} '\n' ;
dataset = '\t' 'dataset' ' ' string '\n' ;
end = 'end' ' ' natural '\n';

scalar-instr = (scalar-summary-instr | ops-instr | 'set' | 'indexvar') ' ' arg [' ' arg] ;
scalar-summary-instr = 'any' | 'all' | 'sum' | 'prod' | 'min' | 'max' ;

matrix-instr = (ops-instr | 'empty' | 'rand' | '%*%') ' ' arg [' ' arg] ;

ops-instr = arith-instr | compare-instr | is-instr | logic-instr | math-instr ;
arith-instr = '+' | '-' | '*' | '/' | '^' | '%%' | '%/%' ;
compare-instr = '==' | '<=' | '>=' | '>' | '<' | '!=' ;
is-instr = 'NA?' | 'INF?' | 'NAN?'
logic-instr = '!' | '|' | '&' ;
math-instr = 'abs' | 'sign' | 'sqrt' | 'floor' | 'ceiling' | 'exp' | 'log' | 'cos' | 'sin' | 'tan' ;

edit-instr = ('update' | 'slice' | 'slice const' | 'dim') ' ' matrix ' ' seq ' ' seq ' ' matrix ;
select-instr = 'select' ' ' arg ' ' arg ' ' arg ;

bind-instr = ('cbind' | 'rbind') ' ' arg ' ' {arg} ;

length = '[1:' natural ']' ;
seq = ordered-seq | unordered-seq ;
ordered-seq = '[' integer ':' integer ':' integer ']' ;
unordered-seq = '[' integer {',' integer} ']' ;

arg = matrix | scalar ;
matrix = '$' natural ;
scalar = pointer | register | value | loop-index;
pointer = '$' natural '@' '(' integer ',' integer ')' ;
register = '%' natural ;
loop-index = '\' natural ;
value = '#' (float | 'NaN') ;

integer = '0' | ['-'] natural ;
natural = nonzero-digit {digit} ;
string = {digit | alpha} ;
float =  digit {digit} '.' digit {digit};
\end{minted}
    \label{fig:dove_grammar}
\end{figure*}
\subsection{Honeybee Genomes and Genomic Evaluation Programs}
\label{app:eval_scripts}

\begin{figure*}
    \centering
    \caption{Absolute runtimes and sizes of the evaluation programs. Programs marked with an * were run on a reduced dataset due to test system limitations. Program \texttt{iES} calls \texttt{EHHS}, so we include the lines of code from \texttt{EHHS} when measuring lines of code for \texttt{iES}. FE are measurements for frontend, NEBE are for measurements with backend without SGX, and EBE are for the backend with SGX. F indicates the use of libFTFP, the data-oblivious floating point arithmetic library that we used on our DOVE implementation. LoC stands for Lines of Code for the original R program whereas DOT size represents the size of the counterpart DOT file in bytes. Finally, the DOT overhead represents the relative overhead of the DOT's file size relative to the size of the original R program. 
    }
    \footnotesize
    \begin{tabular}{l|l|l|l|l|l|l|l|l}
Program & Vanilla R (s) & FE (s) & NEBE (s) & NEBE w/ F (s) &  EBE w/ F (s) & LoC (lines) & DOT size (bytes) & DOT Overhead\\
\hline
$\texttt{EHHS}^{*}$ 
& 18.9&3.85&1104.43&2131.65&3575.46&40&1538&0.51\\
$\texttt{iES}^{*}$
& 23.48&6.43&1106.34&2161.95&3625& 15 + 40&159853&105.44\\
$\texttt{LD}^{*}$ 
& 1787.58&3.64&2869.48&9040&32264&54&5610&0.98\\
\texttt{allele\_sharing} 
& 283.41&5.6&650.03&1841.28&29733& 12&419&0.28\\
\texttt{hwe\_chisq} 
& 38.48&4.56&113.98&262.23&853.49& 21&5295&4.35\\
$\texttt{hwe\_fisher}$ 
& 690.2&4.98&141425&154194&234054& 12&10287&3.92\\
\texttt{neiFis\_multispop} 
& 85.85&16.88&111.82&278.42&1077.44& 38&5311&4.09\\
\texttt{neiFis\_onepop} 
& 39.13&4.9&55.85&192.53&764.38& 19&7381&2.43\\
$\texttt{snp\_stats}$ 
& 692.73&11.21&142783&155840&236644& 33&1980&1.35\\
\texttt{wcFstats} 
& 55.27&8.21&79.38&186.27&757.38& 35&6624&1.58\\
\texttt{wcFst\_spop\_pairs} 
& 74.05&15.43&206.55&458.26&1343.51 & 45&18606&5.21\\
    \end{tabular}
    \label{fig:perf_table}
\end{figure*}

To validate DOVE, we worked with an application of genomic data sharing. The specific application~\cite{bee_study} aims to understand from genomics why honeybees from Puerto Rico are \textit{gentle}, like European honeybees, even though they descend from \textit{aggressive} Africanized honeybees from South America. Genomes from 30 honeybees were collected from Puerto Rico, Mexico, and the United States to provide a total of 90 genomes. 

We reproduce the study in \cite{bee_study} with this data using R and DOVE on SGX, but truncate the total number of samples to 60. 
This study uses one of evaluation programs~\cite{eva_chan} that we explore in Section~\ref{sec: perf_eval: eval_scripts}. 

Overall, this honeybee study simulates the idea that three parties would like to derive critical information from their combined data set without the need for a trusted third party to consolidate the data. We did not work with truly sensitive data in this study, but the characteristics of the data are essentially the same as would have been used in the hypothetical bipolar disorder study mentioned in the introduction. The honeybee data and code is available for download and will be a good benchmark for future studies of this kind.

For our performance evaluation, we run DOVE against all of these evaluation programs. We collect measurements for the frontend separate from the backend, and on three configurations of the backend. More information about the experimental setup can be found at the beginning of Section~\ref{sec: perf_eval}. Figure~\ref{fig:perf_table} shows the absolute measurements that we collected from DOVE in Section~\ref{sec: perf_eval: eval_scripts}. 
Columns FE, NEBE, NEBE w/ F, and EBE, respectively, represent measurements on the frontend, backend without libFTFP and outside of the enclave, backend with libFTFP and outside of the enclave, and backend with libFTFP and inside of the enclave. 
Note that some of these functions were run on first 10,000 rows of our 1.3 GB honeybee dataset (due to testing hardware limitations) while remaining functions were run on the full dataset.
The functionality of each of these programs is explained below.

\begin{compactenum}
    \item \texttt{allele\_sharing} A program to calculate the allele sharing distance between pairs of individuals~\cite{ec_ref1}.
    \item \texttt{EHHS} A program to calculate the EHHS values for a given chromosome~\cite{ec_ref0}.
    \item \texttt{hwe\_chisq} A program to test the significance of deviation from Hardy–Weinberg Equilibrium (HWE) using Pearson's Chi-Squared test.
    \item \texttt{hwe\_fisher} A program to test the significance of deviation from HWE using Fisher's Exact test.
    \item \texttt{iES} A program to calculate the iES statistics~\cite{ec_ref0}. The code calls \texttt{EHHS} in computing its statistics.
    \item \texttt{LD} A program to calculate $D$, $D'$, $r$, ${\chi}^2$, ${\chi}^{2}{'}$, which are statistics based on the frequences of alleles in the input.
    \item \texttt{neiFis\_multispop} A program to calculate inbreeding coefficients, $F_{is}$~\cite{ec_ref2}, for each sub-population from a given set of SNP markers.
    \item \texttt{neiFis\_onepop} A program to calculate inbreeding coefficients, $F_{is}$~\cite{ec_ref2}, for the total population from a given set of SNP markers.
    \item \texttt{snp\_stats} A program to calculate basic stats on SNPs, including: allele frequency, minor allele frequency, and exact estimate of HWE.
    \item \texttt{wcFstats} A program to estimate the variance components and fixation indices~\cite{ec_ref3}.
    \item \texttt{wcFst\_spop\_pairs} A program to estimate $F_{st}$ (\(\theta\)) values for each pair of sub-populations~\cite{ec_ref3}.
\end{compactenum}

\subsection{Experimental Evaluation for Data Obliviousness}
\label{app:pcmapi}
\begin{figure}[t]
    \centering
    \caption{Intel PCM Functions used for dynamic analysis. All function names are in a form of \texttt{getX} where \texttt{X} is the metric that we collect.}
    \footnotesize
    \begin{tabular}{|l|l|}
    \hline
    Function Name & Criterion \\
    \hline
    \texttt{getCycles} & cycle counts\\ 
    \texttt{getAverageFrequency} & frequency\\ 
    \texttt{getActiveAverageFrequency} & frequency\\ 
    \texttt{getCyclesLostDueL3CacheMisses} & cycle counts, cache H/M\\ 
    \texttt{getCyclesLostDueL2CacheMisses} & cycle counts, cache H/M\\ 
    \texttt{getL2CacheHitRatio} & cache H/M\\ 
    \texttt{getL3CacheHitRatio} & cache H/M\\ 
    \texttt{getL3CacheMisses} & cache H/M\\ 
    \texttt{getL2CacheMisses} & cache H/M\\
    \texttt{getL2CacheHits} & cache H/M\\
    \texttt{getL3CacheHitsNoSnoop} & cache H/M\\
    \texttt{getL3CacheHitsSnoop} & cache H/M\\
    \texttt{getL3CacheHits} & cache H/M\\
    \texttt{getBytesReadFromMC} & bytes from/to MC\\
    \texttt{getBytesWrittenToMC} & bytes from/to MC\\
    \texttt{getIORequestBytesFromMC} & bytes from/to MC\\
    \hline
    \end{tabular}
    \label{fig:pcm_funcs}
\end{figure}

\begin{figure}[!htb]
    \centering
    \includegraphics[width=\columnwidth/2]{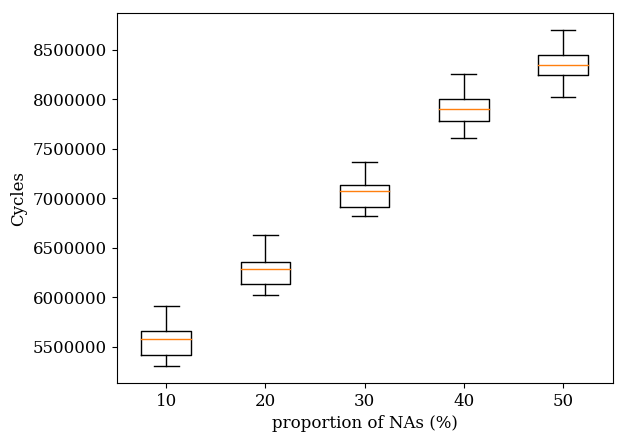}\includegraphics[width=\columnwidth/2]{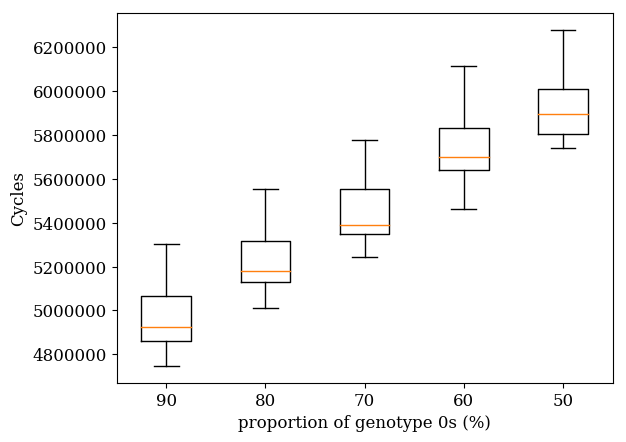}
    \includegraphics[width=\columnwidth/2]{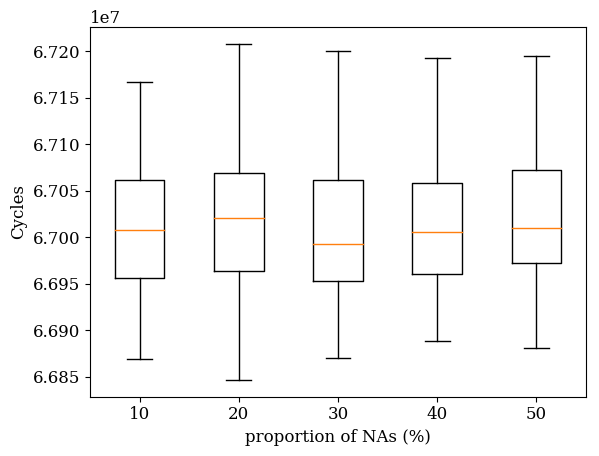}\includegraphics[width=\columnwidth/2]{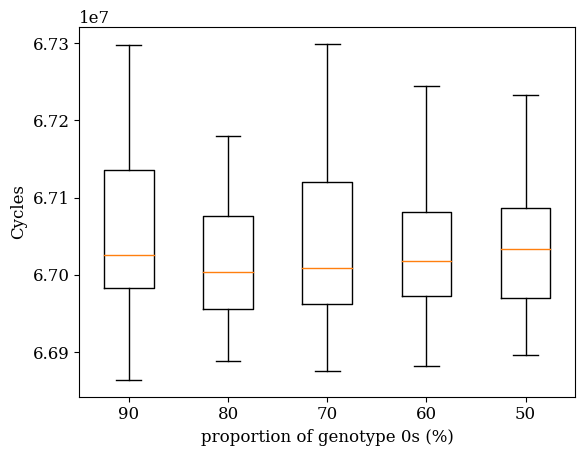}
    \caption{Cycle count measurements for runtime Intel PCM analysis against Line~\ref{line:bwand} of Figure~\ref{fig:r_snippet}. 
    Plots above are measurements from vanilla R and plots below are from DOVE. The plots on the left are tested against varying proportions of \mintinline{s}{NA}, and plots on the right are tested against varying proportions of \mintinline{s}{0}.
    }
    \label{fig:pcm_example}
\end{figure}

We discussed the experimental evaluation of DOVE (Section~\ref{sec: perf_eval}) in three areas: expressiveness, correctness and computational efficiency. In this section, we conduct an experimental evaluation on the data obliviousness of DOVE with the two case studies to supplement our security evaluation in Section~\ref{sec:sec_analysis}, as opcode and execution trace analysis may not be sufficient to cover a diverse array of (and potentially undocumented) $\mu Arch$ side channels stemming from subtle, microarchitecture-specific optimizations.

To illustrate the effectiveness of this measurement-based analysis, we conducted the same experiment against Line~\ref{line:bwand} of Figure~\ref{fig:r_snippet}, the aforementioned \mintinline{s}{&} operator example, on two datasets.
The first set tests whether the function is data oblivious against \mintinline{s}{NA} or not. This set consists of matrices with five different proportions (10\%, 20\%, 30\%, 40\%, 50\%) of \mintinline{s}{NA}.
A second set tests whether the function is data oblivious against \mintinline{s}{0} or not. This set also consists of matrices with five different proportions (90\%, 80\%, 70\%, 60\%, 50\%) of \mintinline{s}{0}. We generate 100 matrices of each proportion randomly in both sets for our testing. 
The size of each matrix is 1,000 by 60. 

Figure~\ref{fig:pcm_example} shows boxplots that illustrate 100 trials of cycle count measurements against the aforementioned two sets of inputs. Each box in the figure represents 100 measurements of random input set with varying proportions of either \mintinline{s}{NA} or \mintinline{s}{0}. 
When Line~\ref{line:bwand} of Figure~\ref{fig:r_snippet} was run on vanilla R, the cycle counts differ drastically when the input's proportion of \mintinline{s}{NA} (top left) or \mintinline{s}{0} (top right) is varied. 
Both plots at the top shows a linear increase in cycle counts as the proportion changes, but measurements from DOVE do not show such trend against \mintinline{s}{NA} (bottom left) or \mintinline{s}{0} (bottom right).

For a more precise analysis, we utilize Intel Processor Counter Monitor (PCM), an API to monitor performance of Intel processors~\cite{intel_pcm}. Among various performance metrics offered by the PCM, some are direct indicators of side-channel vulnerabilities. 
Such $\mu Arch$ measurements include cycle counts and L2/L3 cache hits. We leverage this API to check the data obliviousness of the DOVE function implementations. We use the same two datasets with varying amounts of \mintinline{s}{NA} or \mintinline{s}{0}. 

We examined every performance metric that can be collected from the PCM and chose 16 metrics that (1) are relevant for $\mu Arch$ side-channel detection and (2) do not overlap with the execution trace analysis. 
These metrics cover one or more of four criteria: cycle counts, frequency, cache hits/misses and bytes from/to the memory controller. 
Figure~\ref{fig:pcm_funcs} shows all Intel PCI API functions used for the runtime analysis on the leaf functions of Section~\ref{sec:sec_analysis:leaf_functions}. These API functions all begin with prefix \texttt{get} and are followed by the metric they measure. Most of these functions provide information on one of four criteria, but function \texttt{getCyclesLostDueL3CacheMisses} and \texttt{getCyclesLostDueL2CacheMisses} provide information on both cycle counts and cache hits and misses.

We performed PCM tests on every function in the backend (corresponding to functions of the Safe and Unsafe DOT/Core groups in Figure~\ref{fig:implemented_function}) on all 16 metrics. We confirm that all DOVE functions are data oblivious against all metrics, regardless of the data with varying proportions of missing data or zeros. 

\fi 

\end{document}